\documentclass[aps,prl,twocolumn,showpacs,psfig,superscriptaddress,longbibliography]{revtex4-1}

\usepackage{textcomp}
\usepackage{times}
\usepackage{graphicx}
\usepackage{float}
\usepackage{latexsym,amsmath,amssymb,bm,euscript}
\usepackage{color}
\usepackage{subfigure}
\usepackage{epstopdf}
\usepackage[colorlinks=true,linkcolor=blue,citecolor=blue,urlcolor=blue]{hyperref}
\usepackage{hyperref}
\usepackage{soul}
\usepackage[normalem]{ulem}
\usepackage{mathrsfs}
\usepackage{amsmath,amssymb}
\DeclareMathOperator*{\argmin}{arg\,min}
\usepackage{lettrine}
\usepackage{xfrac}
\usepackage{xspace}
\usepackage{textcomp}
\usepackage{wasysym}
\usepackage{xr}
\newcommand{\Eq}[1]{Eq.~\eqref{#1}}

\newcommand{\Fig}[1]{Fig.~\ref{#1}}

\newcommand{\highlight}[1]{{\color[rgb]{0,0,0}{#1}}}

\graphicspath{{./Figs/}}

\begin{document}
\title{Ubiquitous nematic Dirac semimetal emerging from interacting quadratic band touching system}
\author{Hongyu Lu}
\affiliation{Department of Physics and HKU-UCAS Joint Institute
	of Theoretical and Computational Physics, The University of Hong Kong,
	Pokfulam Road, Hong Kong SAR, China}
\author{Kai Sun}
\email{sunkai@umich.edu}
\affiliation{Department of Physics, University of Michigan, Ann Arbor, Michigan 48109, USA}
\author{Zi Yang Meng}
\email{zymeng@hku.hk}
\affiliation{Department of Physics and HKU-UCAS Joint Institute
	of Theoretical and Computational Physics, The University of Hong Kong,
	Pokfulam Road, Hong Kong SAR, China}
\author{Bin-Bin Chen}
\email{bchenhku@hku.hk}
\affiliation{Department of Physics and HKU-UCAS Joint Institute
	of Theoretical and Computational Physics, The University of Hong Kong,
	Pokfulam Road, Hong Kong SAR, China}

\begin{abstract}

Quadratic band touching (QBT) points are widely observed in 2D and 3D materials, including bilayer graphene and Luttinger semimetals, and attract significant attention from theory to experiment. However, even in its simplest form, the 2D checkerboard lattice QBT model, the phase diagram characterized by temperature and interaction strength still remains unknown beyond the weak-coupling regime. Intense debates persist regarding the existence of various interaction-driven insulating states in this system~\cite{KSun2009,SUebelacker2011_qbt,HQWu2016_qah,SSur2018_QBT,TSZeng2018_QBT,HYLu2022_qbt,xidai2022}. To address these uncertainties, we employ thermal tensor network simulations, specifically exponential tensor renormalization group and tangent space tensor renormalization group~\cite{BBChen2018_XTRG,tanTRG2023}, along with density matrix renormalization group calculations to provide \highlight{a comprehensive finite-temperature phase diagram for this model and shed light on previous ambiguities. Notably, our findings reveal the emergence of a robust bond-nematic Dirac semimetal phase with distinct thermodynamic properties that set it part from the nematic insulating state and other symmetry broken states. This previously overlooked feature is found to be ubiquitous in interacting QBT systems.} We also discuss the implications of these results for experimental systems such as bilayer graphene and iridate compounds.
\end{abstract}

\date{\today }
\maketitle

\noindent{\textcolor{blue}{\it Introduction.}---} 
Quadratic band touching (QBT) points are widely observed in materials. In the 2D realm, bilayer graphene serves as a representative example~\cite{yanFormation2011,langAntiferromagnetism2012,SPujari2016_graphene,Lukas2018_qbt,Lukas2021_qbt}, while in 3D, materials like $\alpha$-Sn, HgTe, and pyrochlore iridates $R_2$Ir$_2$O$_7$ ($R=$ Pr, Nd) exhibit Luttinger semimetal behavior~\cite{luttingerQuantum1956,moonNonFermi2013,herbutTopological2014,chengDielectric2017,wangUnconventional2020,kondoQuadratic2015,boettcherAnisotropy2017,janssenPhase2017,kharitonovEvolution2022}. Investigating the effects of interactions on QBTs is both fundamentally significant and experimentally relevant. For instance, understanding how QBTs transform into symmetry-breaking phases in bilayer graphene under the influence of trigonal warping and Coulomb interactions, or how Luttinger semimetals can evolve into Dirac, line-{nodal}, or Weyl semimetals, and even non-Fermi-liquid states in the presence of strain, bulk-inversion asymmetry, and Coulomb interactions, remains an active area of intense theoretical and experimental investigations~\cite{yanFormation2011,langAntiferromagnetism2012,SPujari2016_graphene,Lukas2018_qbt,Lukas2021_qbt,moonNonFermi2013,herbutTopological2014,chengDielectric2017,wangUnconventional2020,kondoQuadratic2015,boettcherAnisotropy2017,janssenPhase2017,kharitonovEvolution2022}.

\begin{figure}[htp!]
	\centering
	\includegraphics[width=\columnwidth]{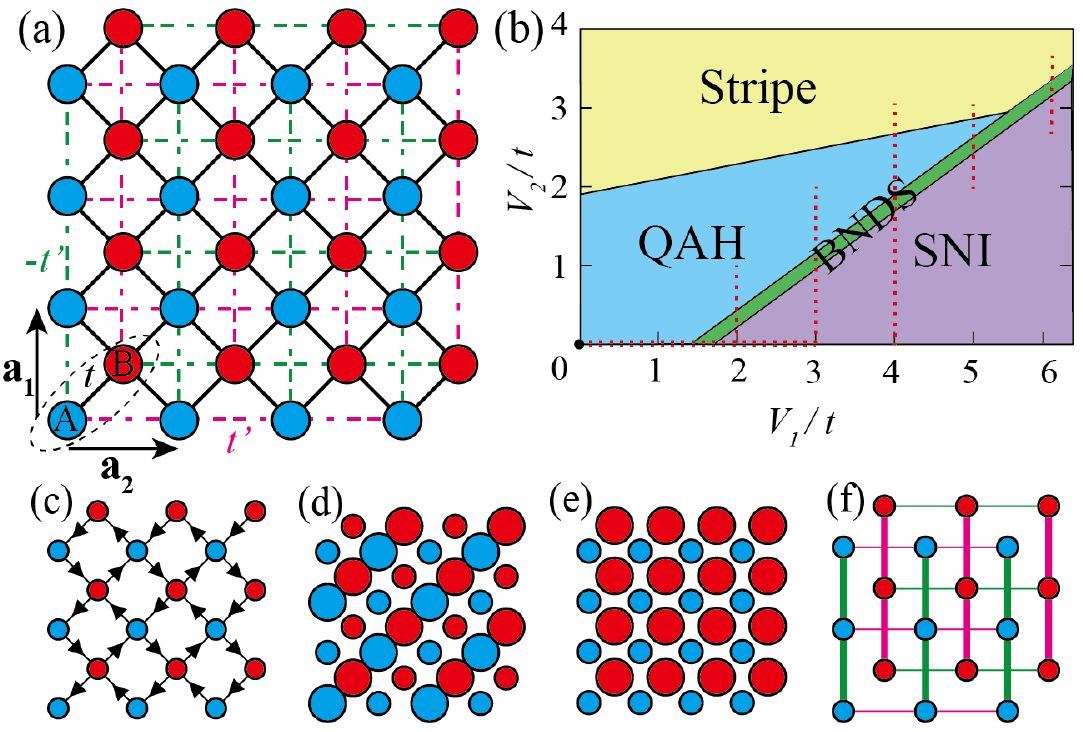}
	\caption{
	(a) The checkerboard lattice with blue (red) dots denoting the A (B) sublattices. 
(b) The comprehensive ground-state phase diagram. 
The QBT is denoted by the black dot at the origin, while the red dotted lines represent the paths of simulations.
(c)-(f) 
The real-space signature of the QAH, Stripe, SNI, and BNDS phases, respectively. The QAH phase is depicted with arrows indicating the loop currents. In the Stripe and SNI phases, large (small) dots represent high (low) electron density. In the BNDS phase, the thickness of the bonds corresponds to the absolute value of the hopping amplitude.}
	\label{fig_lattice_phase_diagram}
\end{figure}

The 2D checkerboard lattice model with $C_4$ symmetry serves as the simplest platform to investigate the effects of interactions on QBTs. Previous studies have revealed that a quantum anomalous Hall (QAH) state with spontaneously broken time-reversal symmetry (TRS) \cite{DJThouless1982, naturereview_QAH, Haldane1988, changExperimental2013, dengQuantum2020, stepanovCompeting2021, haldaneBerry2004, Nagaosa2010, DiXiao2010,SZZhang2008_TMI,BBChen2021_TBG,JPLiu2021_tbgqah,XYLin2022_exciton, GPPan2023_qah,zhangQuantum2022,linExciton2022,huangEvolution2023} emerges in the weak-coupling regime~\cite{KSun2009, SUebelacker2011_qbt, HQWu2016_qah, SSur2018_QBT,  TSZeng2018_QBT, HYLu2022_qbt}.
Proposals have also been put forth to realize QBT and interaction-driven QAH states in real materials, which include the four-band checkerboard lattice system based on monolayer CrCl$_2$(pyrazine)$_2$~\cite{xidai2022}, as well as the $C_6$ symmetric Kagome metals~\cite{SNishimoto2010,JWen2010,WZhu2016_qah,HYHui2018,sankarObservation2023,YFRen2018_qah}.

Beyond the weak-couling regime, the QAH phase is anticipated to give way to other competing phases, such as the $C_2$ symmetric site-nematic insulator SNI~\cite{KSun2009}. However, due to the strong coupling nature of these phases, a complete phase diagram 
remains a subject of debate, with conflicting results~\cite{KSun2009, SUebelacker2011_qbt, HQWu2016_qah, SSur2018_QBT,  TSZeng2018_QBT, HYLu2022_qbt,xidai2022}. 
Exact diagonalization (ED) and density matrix renormalization group (DMRG) simulations suggest 
that there is no co-existing region between the QAH and SNI states~\cite{HQWu2016_qah}. Instead, a $C_4$ symmetry-breaking bond-nematic Dirac semimetal (BNDS) phase emerges as an intermediate phase, where the QBT splits into two Dirac cones~\cite{TSZeng2018_QBT}. Notably, the BNDS and SNI were identified as distinct thermodynamic phases with different order parameters~\cite{TSZeng2018_QBT}.

The transitions between QAH, BNDS, and SNI phases are found to be first-order, 
different from the interaction-driven phase transitions in Bernal stacking graphene~\cite{langAntiferromagnetism2012,SPujari2016_graphene,Lukas2018_qbt,Lukas2021_qbt}. And the previous studies~\cite{KSun2009, SUebelacker2011_qbt, HQWu2016_qah,  TSZeng2018_QBT} considered only nearest-neighbor (NN) interactions 
without next-nearest-neighbor (NNN) interactions, which are crucial for understanding the transition to other interaction-driven states such as the stripe-like charge density wave insulator (Stripe). Furthermore, the relationship between the BNDS and SNI phases has not been adequately addressed, and we demonstrate that they break exactly the same symmetry.

By incorporating NNN interactions into the system, 
Ref.~\cite{SSur2018_QBT} observed a direct transition from QAH or Stripe to the SNI phase, omitting the BNDS phase. In a Hartree-Fock study of a doubled four-band checkerboard lattice~\cite{xidai2022}, the BNDS phase 
was not reported. Instead, they discovered a Dirac semimetal phase with site-nematic order between the site-nematic insulator and the QBT semimetal. These discrepancies highlight the ongoing challenges in achieving a consensus on the ground state phase diagram of the checkerboard lattice QBT system. 
\highlight{Furthermore, the finite-temperature phase diagram, especially the distinct $C_V \sim T^2$ behavior from emergent Dirac cones of the interaction-driven BNDS phase, remains completely unexplored.}

In this Letter,
\highlight{by employing the exponential and tangent space tensor renormalization groups (XTRG and {\it tan}TRG)~\cite{BBChen2018_XTRG,tanTRG2023} and DMRG calculations, we unveil a comprehensive phase diagram based on thermodynamic characteristics. One key finding is the ubiquitous presence of the BNDS phase, which separates not only the QAH and SNI phases, but also the Stripe and SNI phases. We clarify the equivalence of SNI and BNDS in terms of symmetry and reveal their distinct measurable properties, especially in their different thermodynamic properties}. We foresee this overlooked interaction-driven BNDS with emergent Lorentz symmetry and unique thermodynamic properties could be a universal feature in interacting QBT systems, extending beyond the checkerboard lattice, with potential realizations in bilayer graphene~\cite{langAntiferromagnetism2012,SPujari2016_graphene,Lukas2018_qbt,Lukas2021_qbt}, $\alpha$-Sn, HgTe and iridate compounds~\cite{luttingerQuantum1956,moonNonFermi2013,herbutTopological2014,chengDielectric2017,wangUnconventional2020,kondoQuadratic2015,boettcherAnisotropy2017,janssenPhase2017,kharitonovEvolution2022}.

\begin{figure*}[htp!]
	\centering	
	\includegraphics[width=\textwidth]{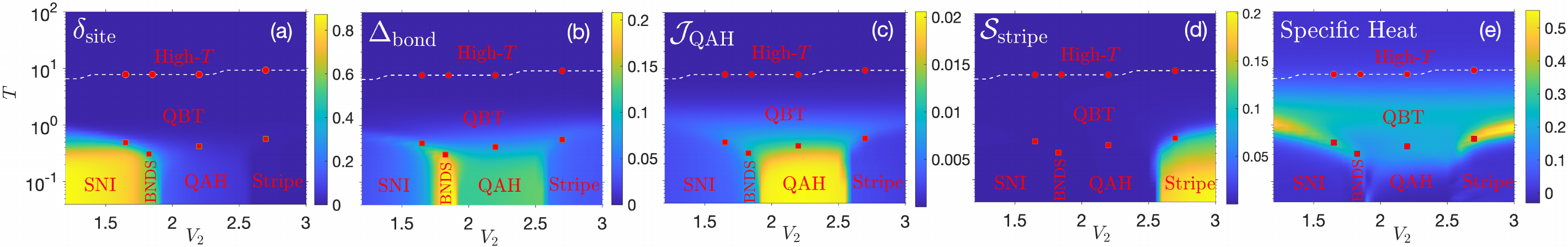}
	\caption{The $T-V_2$ phase diagrams at $V_1=4$. 
		Plots show (a) the SNI order parameter, (b) BNDS order parameter, (c) QAH structure factor, (d) stripe structure factor, and (e) specific heat $c_V$.
		The white dashed lines indicate the boundaries between the high-temperature thermal metallic regime and the intermediate-temperature QBT regime, identified using the red circles, which represent the crossover temperature between the high-$T$  and QBT regimes, as illustrated in Figure \ref{cutfig_v1_4}(a1-d1). 
		The red squares denote the transition temperature between the QBT regime to the low-temperature symmetry-breaking phases.}
	\label{colorfig_v1_4}
\end{figure*}

\noindent{\textcolor{blue}{\it Model, Method and Ground-State Phase Diagram.}---} We examine a spinless fermion model in a checkerboard lattice,
\begin{equation}
\begin{aligned}
H = \sum_{\mathbf{r}}[ & t~\sum_{\gamma}c^\dag_{\mathbf{r},B} c^{\phantom{\dag}}_{\mathbf{r+\gamma},A} + \mathrm{h.c.} &\text{ (NN)}\\ 
+& t' \sum_{\lambda,i} (\eta^{\phantom{\dag}}_\lambda \epsilon^{\phantom{\dag}}_i) c^\dag_{\mathbf{r},\lambda} c^{\phantom{\dag}}_{\mathbf{r+a_i},\lambda}  + \mathrm{h.c.} &\text{ (NNN)} \\
+ & V_1\sum_{\gamma}(n_{\mathbf{r},B}-\tfrac{1}{2}) (n_{\mathbf{r+\gamma},A}-\tfrac{1}{2}) &\text{ (NN)}\\ 
+& V_2\sum_{\lambda,i} (n_{\mathbf{r},\lambda}-\tfrac{1}{2}) (n_{\mathbf{r+a_i},\lambda}-\tfrac{1}{2}) ]\text{,} &\text{ (NNN)}\label{hamiltonian}
\end{aligned}
\end{equation} 
with $\gamma\in\{\mathbf{0},\mathbf{a_1},\mathbf{a_2},\mathbf{a_1+a_2}\}$, $\lambda\in\{A,B\}$ being the sublattice index, 
$\{\eta_A=1, \eta_B=-1\}$ and $\{\epsilon_1=1, \epsilon_2=-1\}$, as shown in Fig.~\ref{fig_lattice_phase_diagram}(a). We fix $t=-1$ and $t'=-0.5$ and consider both the repulsive NN and NNN density-density interactions.

We mainly employ the XTRG and {\it tan}TRG method~\cite{BBChen2018_XTRG,Li2019_XTRGBenchmark,tanTRG2023} for finite-temperature simulations
 \cite{Chen2019_TwoTemperature,Li2019_XTRGBenchmark,Li2020_KT,Chen2021_ColdAtom,XYLin2022_exciton,Qu2022_tJ}
with YC$4\hspace{-0.3em}\times\hspace{-0.3em}16$ geometry and 
we retain up to $D=1000$ bond states, rendering the truncation errors $\delta < 10^{-4}$ down to the low-$T$ regime. We also perform YC$6$ DMRG simulations, with up to $4096$ bond states and maximal truncation errors $\delta \approx 10^{-5} $. 

\begin{figure*}[htp!]
\centering
\includegraphics[width=\textwidth]{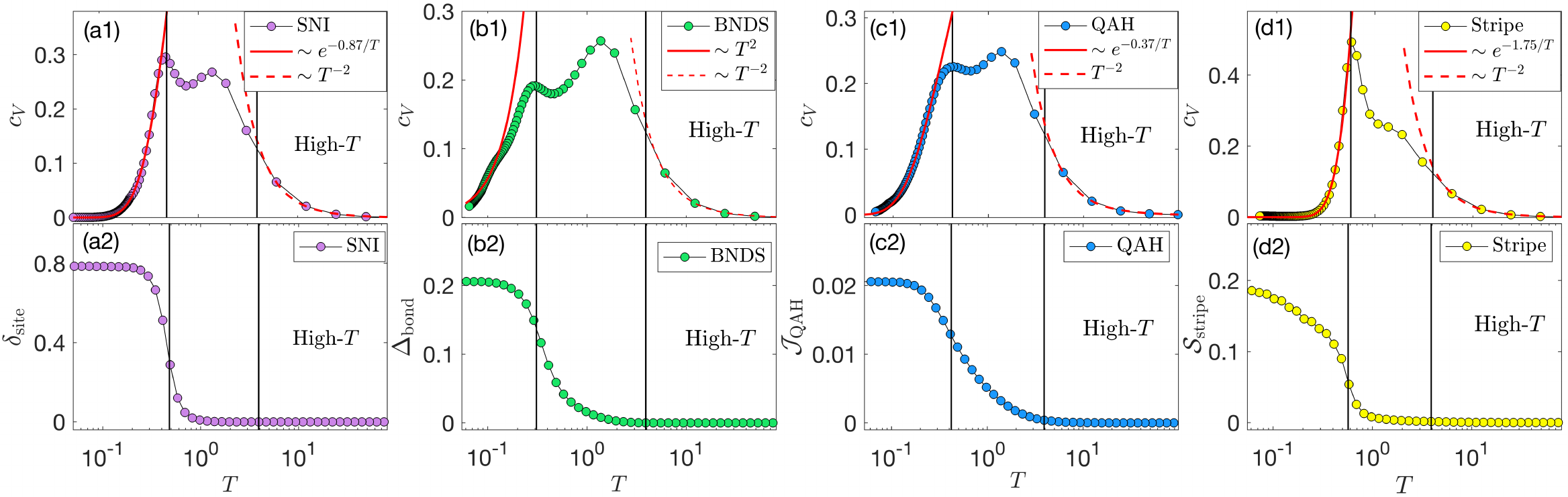}
\caption{The thermodynamic characteristics along the cuts in Fig.~\ref{colorfig_v1_4}  as a function of temperature $T$. With fixed $V_1=4$, 
(a1, a2) represent SNI at $V_2=1.65$, (b1, b2) represent BNDS at $V_2=1.85$,  (c1, c2) represent QAH at $V_2=2.2$, (d1, d2) represent Stripe at $V_2=2.7$. The first row displays the specific heat, while the second row shows the order parameters or structure factors for each phase. 
The separation temperature between the low-temperature, intermediate, and high-temperature regimes is indicated by the black solid lines. \highlight{The $C_V\sim T^2$ for BNDS at low temperature is distinctively different from the $C_V \sim \exp(-\Delta/T)$ for the three other phases.} }
\label{cutfig_v1_4}
\end{figure*}

Here we set $V_1=4$ and systematically vary $V_2$ to demonstrate the existence of the BNDS phase between the QAH and SNI states. Additional numerical evidence supporting the presence of BNDS phases for different $V_1$ can be found in the Supplemental Materials (SM)~\cite{suppl}.
Furthermore, we investigate the large-interaction regime by fixing $V_1=6$ and varying $V_2$, revealing the occurrence of the BNDS phase between the Stripe and SNI phases. \highlight{The BNDS state is unambiously characterized from both the low-$T$ 
power law behaviour of specific heat ($C_V \sim T^2$), and the changes of conformal central charge $c$ 
and the single-particle excitation gap via flux insertion.}
Consequently, we construct a comprehensive ground-state phase diagram, as illustrated in 
Fig.~\ref{fig_lattice_phase_diagram}(b). 
\highlight{Additionally, through comparing the symmetry of the two nematic states, BNDS and SNI, we find both states break exactly the same symmetry. In other words, akin to the gas-liquid and/or Lifshitz transitions, the first-order phase boundary between them does not involve a change in symmetry. }

it is possible for the establishment of an adiabatic pathway between these two states through exploring a higher-dimensional phase space. Regarding the bond- and site-nematic order parameters, without fine-tuning, both of these order parameters remain nonzero in both the BNDS and SNI phases, reflecting their identical symmetry. This finding contrasts with the numerical results obtained from finite-size extrapolation as reported in previous work~\cite{TSZeng2018_QBT}. 
\noindent{\textcolor{blue}{\it $T-V$ phase diagram and the thermodynamic characteristics.}---} 
To identify different phases in the $T-V$ phase diagram, we utilize several measurements: {\it QAH structure factor} $\mathcal{J}_\mathrm{QAH}\equiv\frac{1}{N_b}\sum_{\langle i,j\rangle}\epsilon_{i,j}\langle\mathcal{C}_{ij}\mathcal{C}_{i_0j_0}\rangle$,  where the summation is over NN bonds in the bulk with $N_b$ being the total number of bonds and $\epsilon_{i,j}=\pm 1$ charactering the orientation of the current $\mathcal{C}_{ij}\equiv (c_i^\dagger c^{\,}_j -c_j^\dagger c^{\,}_i )$; 
{\it bond nematic order parameter} $\Delta_\mathrm{bond}\equiv|\langle c^\dag_{\mathbf{r},\lambda} c^{\phantom{\dag}}_{\mathbf{r+a_2},\lambda}\rangle|-|\langle c^\dag_{\mathbf{r},\lambda} c^{\phantom{\dag}}_{\mathbf{r+a_1},\lambda}\rangle|$, which measures the difference in NNN hopping amplitudes along the vertical ($\mathbf{a_2}$) and horizontal ($\mathbf{a_1}$) directions;
{\it site nematic order parameter} $\delta_\mathrm{site}\equiv\langle n_\mathrm{B}-n_\mathrm{A}\rangle$, 
which quantifies the difference of  electron density at `A' and `B' sublattices;
{\it structure factor for the Stripe phase} $\mathcal{S}_\mathrm{stripe} \equiv \frac{1}{N}\sum_{i} e^{-i\vec{k}\cdot\vec{r}_{0,i}} (\langle \hat{n}_i\hat{n}_0 \rangle -\langle\hat{n}_i\rangle \langle\hat{n}_0\rangle )$,
where $\vec{k}=(\pm\frac{\sqrt{2}\pi}{2},\mp\frac{\sqrt{2}\pi}{2})$ represents the Stripe pattern shown
in \Fig{fig_lattice_phase_diagram}(d);
{\it the specific heat} $c_V(T)\equiv\tfrac{\partial E(T)}{\partial{T}}$, where $E$ is the energy density.

In Fig.~\ref{colorfig_v1_4}(a-d), we present various order parameters and structure factors respectively. At $V_1=4$, when $V_2$ is small (Fig.~\ref{colorfig_v1_4}(a)), the ground state is a SNI, and the order parameter $\delta_{\mathrm{site}}$, as well as the transition temperature, reduces as $V_2$ increases.
In Fig.~\ref{colorfig_v1_4}(b), an intermediate BNDS phase between QAH and SNI is clearly observed. This phase exhibits a strong bond nematic order $\Delta_{\mathrm{bond}}$ 
Notably, the site nematic order $\delta_{\mathrm{site}}$ is also present in the BNDS region and becomes stronger as $V_2$ decreases, ultimately exhibiting a significant jump across the phase boundary to the SNI phase. Furthermore, a sharp change in the QAH (Fig.\ref{colorfig_v1_4}(c)) and Stripe   (Fig.\ref{colorfig_v1_4}(d))  structure factors near the phase boundary $V_2\sim2.6$ indicates a first-order transition between these two phases. The specific heat shown in Fig.~\ref{colorfig_v1_4}(e) also supports the identification of these phases with distinct characteristics.

To analyze the detailed thermodynamic transitions and features, we present four cuts of the 2D phase diagrams in Fig.~\ref{cutfig_v1_4}. These cuts are taken along the temperature axis at fixed interactions, corresponding to four different ground states. 
At high temperatures (much larger than the band width), the specific heat exhibits a scaling behavior of $c_V\sim T^{-2}$ for all cases.
A crossover temperature can be identified where $c_V$ deviates from the $T^{-2}$ scaling, 
below which, $c_V$ displays two peaks or shoulders. The lower-temperature one, 
marks the transition to the low-$T$ ground state, which agrees with the order parameters / structure factors versus temperature
, while the higher-temperature peak/shoulder is a thermodynamic characteristic of the electron band structure.
\highlight{With the latest {\it tan}TRG method, we are capable of fitting the low-$T$ behavior of $c_V$, 
which has the exponential activation of $C_V \sim \exp(-\Delta/T)$ in QAH, SNT and Stripe phases. While more importantly, we can see the $T^2$ behavior of low-$T$ $c_V$ in BNDS phase, 
which is the expected behavior of 2D Dirac cones and further supports the BNDS phase is actually a Dirac semimetal, emerging from interaction-driven QBT systems.}

We note that for the thermodynamics of QAH state, at $V_1=4$, both its transition temperature and the amplitude of the loop current exhibit little dependence on $V_2$, as illustrated in Fig.~\ref{colorfig_v1_4}(c) and SM~\cite{suppl}.
In contrast, when $V_2$ is set to zero, the transition temperature and the amplitude of the loop current of the QAH state exhibit a strong dependence on $V_1$. As $V_1$ increases, both these two quantities increase before saturating at certain constant values. Besides, we add the DMRG results of QAH order parameters to SM~\cite{suppl} for more evidence of QAH state, which agrees with the finite-temperature phase diagram.

\begin{figure}[htp!]
	\centering
	\includegraphics[width=\columnwidth]{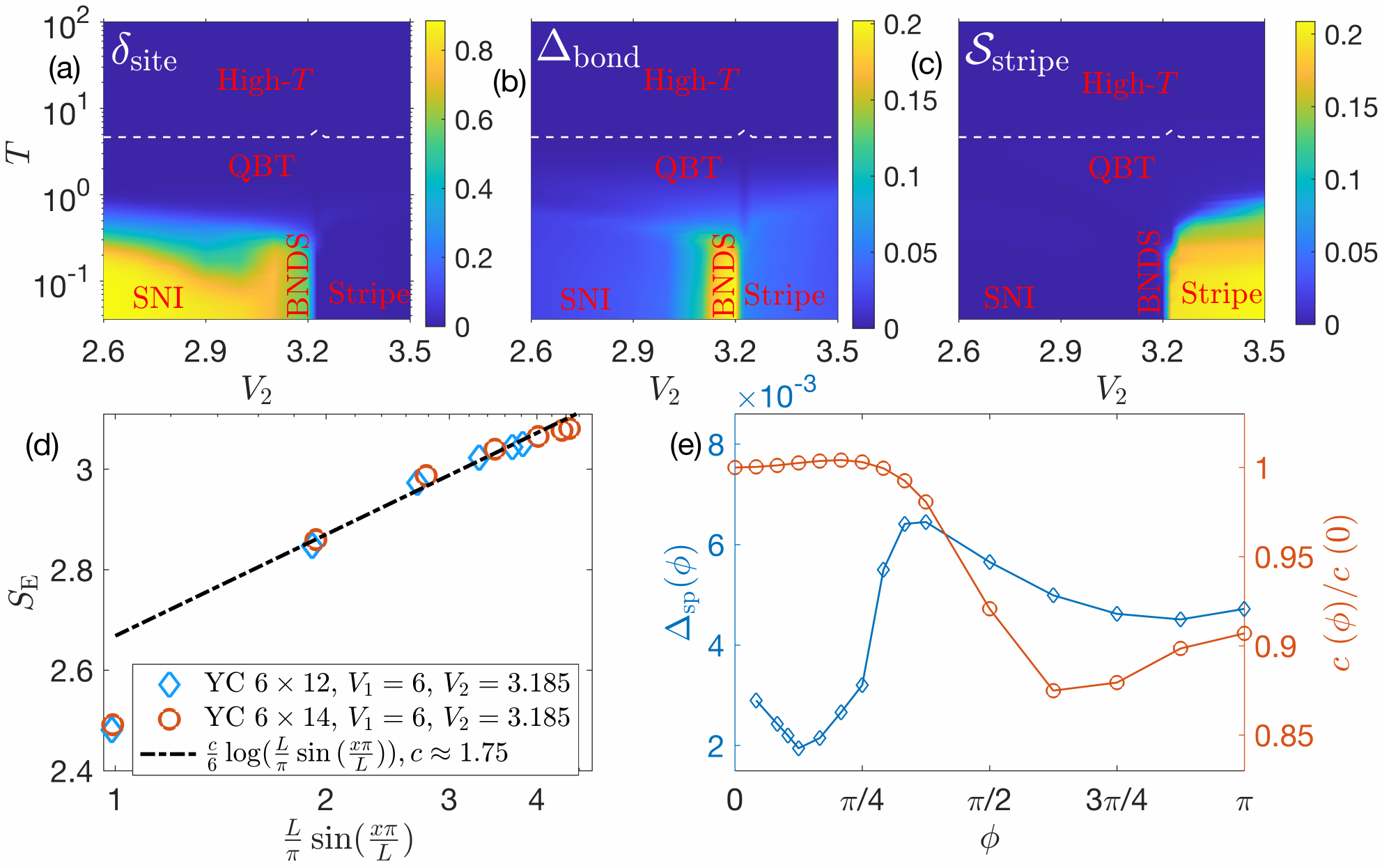}
	\caption{The $T-V_2$ phase diagram with $V_1=6$, including (a) the site-nematic order, (b) the NNN bond nematic order, and (c) the structure factor for stripe. 
		(d) Entanglement entropy in BNDS phase, with log-scale horizontal axis. 
		\highlight{(e) With twisted boundary conditions in BNDS, by inserting flux $\phi$, the single-particle gap and fitted central charge compared with that of no flux are shown. Corresponding shift of momenta grids is $\phi / L_y$.}}
	\label{color_v1_6}
\end{figure}

More importantly, with larger interactions, the QAH phase disappears, but the BNDS always survives as an intermediate phase between SNI and Stripe, in direct contrast to the predictions from all existing studies. 
The color figures in Fig.\ref{color_v1_6}(a-c), clearly depict the Stripe-BNDS-SNI transition, with fixed $V_1=6$ and varying $V_2$. To further verify the presence of the intermediate BNDS phase, we perform DMRG simulations within the BNDS phase and extract the conformal central charge from the entanglement entropy. 
By fitting the entanglement entropy as a function of the cylinder length to the universal scaling function $S_{\mathrm{E}}(x)=\frac{c}{6}\log(\frac{L}{\pi}\sin{\frac{x\pi}{L}})$, we obtain a central charge $c\approx 1.75$ with $D=4096$ (which is still far from convergence), as shown in Fig.~\ref{color_v1_6}(d) and SM\cite{suppl}, which directly supports the existence of Dirac cones with emergent Lorentz symmetry in the BNDS phase. \highlight{To further check the location of Dirac cones, we implement the twisted boundary conditions in the $\hat{y}$ direction by inserting the flux 
 , $c_i^\dagger c_j^{\ }+h.c. \rightarrow c_i^\dagger c_j^{\ }e^{i\phi}+h.c.$, for all hopping terms across the boundary.
 For the same parameters in BNDS phase, with $V_1=6$ and $V_2=3.185$, 
 we fit the central charge $c$ with certain flux $\phi$, compared with that of no flux, and also measure the single-particle gap $\Delta_{\mathrm{sp}}(\phi)$ by extrapolating the ground-state energy at half-filling sector and lowest energy in the single-particle-excited sector (details in SM\cite{suppl}).
 As shown in Fig.\ref{color_v1_6}(e),
 It is clear that the positions of Dirac cone are close to $(\pi,\pi \pm \pi/3)$ with a minute shift approximated $\phi/L_y\pm\pi/48$ (as denoted by the minimal of the single-particle gap, where the fitted central charge also saturates at its peak value), which is consistent with the meanfield analysis in the SM~\cite{suppl}, and in this sense, the YC6 cylinder geometry is already close to be able to host the Dirac cones in the BNDS phase and our evidence of the Dirac cones are robust.}

\begin{figure}[htp!]
	\centering
	\includegraphics[width=\columnwidth]{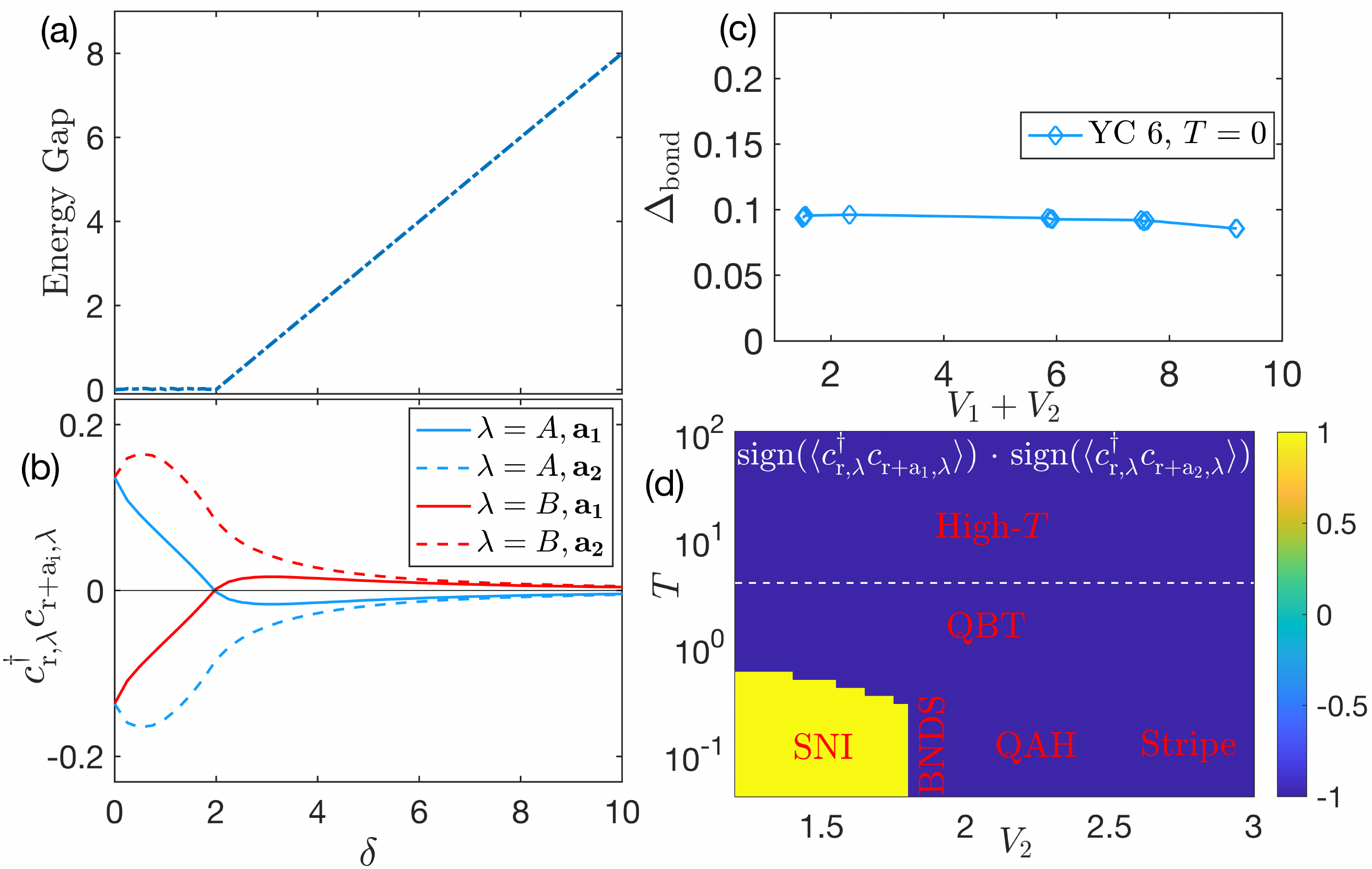}
	\caption{(a-b) Meanfield analysis. With increasing site-nematic order $\delta$, the opening of energy gap is shown in (a) and expectation values of NNN hoppings along $\mathbf{a_1}$ ($\mathbf{a_2}$) directions for two sublattices respectively are shown in (b). (c) DMRG results of the NNN bond order of the BNDS phase throughout the phase diagram.
		(d) The product of signs of NNN hoppings along two directions with fixed $V_1=4$, which is $\pm1$ if the hoppings take the same (opposite) sign(s). }
	\label{mf_figs}
\end{figure}

Furthermore, it is noteworthy that the strength of the NNN bond order in different regions of the BNDS phase remains almost constant around 0.1, as depicted in Fig.~\ref{mf_figs}(c).
Combining simulations at different $V_1$ and $V_2$, the phase diagram shown in Fig.~\ref{fig_lattice_phase_diagram}(b) indicate that this BNDS arises universally as an intermediate phase between SNI and all other grounds states in the phase space.

\noindent{\textcolor{blue}{\it Mechanism of phase transition and nematic phases.}---} We further perform mean-field analysis to better understand the transitions and the bond (site) nematic phases. For the BNDS, we change the NNN hopping $t'\to (t'-\epsilon_i\Delta_{\mathrm{bond}})$ in \Eq{hamiltonian} with positive $\Delta_{\mathrm{bond}}$;  
for the SNI, we add sublattice chemical potential 
$\delta_{\mathrm{site}} \sum_\mathbf{r}(n_{\mathbf{r},A} - n_{\mathbf{r},B})$
to \Eq{hamiltonian}.
At the quadratic level, a finite value of both $\delta_{\mathrm{site}}$ and $\Delta_{\mathrm{bond}}$ 
will split the QBT at ($\pi,\pi$) into two Dirac cones. 
From the perspective of symmetries, the QAH state breaks the TRS, while the BNDS and SNI states break the same lattice $C_4$ rotation symmetry into the same $C_2$ symmetry. 
In fact, our mean-field calculation (c.f. SM~\cite{suppl}) shows, 
the site (bond) nematic term will also induce bond (site) order, which is consistent with our numerical results and clarifies the confusion in previous work~\cite{TSZeng2018_QBT}, that the two nematic orders should be the same. 

The difference between SNI and BNDS is not in symmetry but in their physical properties including, 
1) with sufficiently strong site-nematic order $\delta_{\mathrm{site}}$, the two Dirac cones will merge and gap out as shown in \Fig{mf_figs}(a), 
whereas the bond-nematic $\Delta_{\mathrm{bond}}$ term cannot gap out the Dirac cone;
2) right after the gap-open point, the sign structure of the NNN hopping amplitudes changes and 
the signs within the same sublattice become the same as shown in \Fig{mf_figs}(b). 
We thus utilize such a feature to distinguish the SNI with BNDS phase. Such as in the $T$-$V$ phase diagram \Fig{mf_figs}(d), the sign structure of NNN hoppings clearly gives the SNI-BNDS boundary. 

In our mean-field analysis, with the increase of $\delta_{\mathrm{site}}$, the two Dirac cones gradually merge into each other at the periodic boundary, and gap out the system, corresponding to the SNI observed.
In reality, between the BNDS and SNI phases is an abrupt first-order transition, in that, although the two splited Dirac cones move away from the ($\pi$,$\pi$) point, they cannot gradually merge into each other in BNDS states but suddenly jump to SNI state with much larger $\delta_{\mathrm{site}}$ around the first order transition, and the system is gapped out. Since there is no symmetry difference, the BNDS-SNI transition is similar to the liquid-gas transition, and one can in principle turn the first-order phase transition to an adiabatic crossover or a critical point through paths in higher-dimensional parameter space.

\noindent{\textcolor{blue}{\it Discussions.}---}In conlusion, we find the interaction-driven BNDS ubiquitously exist beween SNI and the other interaction-driven insulating states in the checkerboard lattice QBT system, and we provide the complete ground state and finite-temperature phase diagram with repulsive interactions, which solves the previous debates in this system~\cite{KSun2009,SUebelacker2011_qbt,HQWu2016_qah,SSur2018_QBT,TSZeng2018_QBT,HYLu2022_qbt,xidai2022}. We find the transition between BNDS and SNI in the phase diagram is similar to the liquid-gas transition and provide the numerical results of the thermodynamic characteristics, \highlight{in that the BNDS state is characterized by $C_V \sim T^2$ of emergent Dirac cones from interacting QBT systems, distinctively different from the exponential form in other phases. }The interaction-driven BNDS, with emergent Lorentz symmetry and unique thermodynamic properties, can stimulate deeper understanding of the QBT systems and provide a new alternative interaction-driven Dirac semimetal in materials such as Kagome metals~\cite{SNishimoto2010,JWen2010,WZhu2016_qah,HYHui2018,YFRen2018_qah,sankarObservation2023}, bilayer graphene~\cite{langAntiferromagnetism2012,SPujari2016_graphene,Lukas2018_qbt,Lukas2021_qbt} and the $\alpha$-Sn, HgTe and iridate compound $R_2$Ir$_2$O$_7$ ($R=$ Pr, Nd)~\cite{wangUnconventional2020,kondoQuadratic2015,boettcherAnisotropy2017,janssenPhase2017}.

\begin{acknowledgments}
	{\it Acknowledgments}\,---\,
	We thank Wei Li and Michael Scherer for discussions, and we in particular thank Lukas Janssen for the enlightening introduction of Luttinger semimetals and discussion on the interaction effect on QBT. HYL thank Shou-Shu Gong for helpful discussions. HYL, BBC and ZYM acknowledge the support from the Research Grants Council (RGC) of Hong Kong SAR of China (Project Nos. 17301420, 17301721, AoE/P-701/20, 17309822,  HKU C7037-22G), the ANR/RGC Joint Research Scheme sponsored by RGC of Hong Kong and French National Research Agency (Project No. A\_HKU703/22), the K. C. Wong Education Foundation (Grant No. GJTD-2020-01). We thank HPC2021 system under the Information Technology Services and the Blackbody HPC system at the Department of Physics, University of Hong Kong for providing computational resources that have contributed to the research results in this paper.
\end{acknowledgments}

%


\newpage\clearpage
\renewcommand{\theequation}{S\arabic{equation}} \renewcommand{\thefigure}{S%
	\arabic{figure}} \setcounter{equation}{0} \setcounter{figure}{0}
\begin{widetext}
	
	\section{Supplemental Materials for \\[0.5em]
		Ubiquitous nematic Dirac semimetal emerging from interacting quadratic band touching system}
	In Supplementary Materials Sec. I, we briefly summarized  exponential tensor renormalization group method and tangent space tensor renormalization group method, and its measurements on thermodynamic quantities.  In Supplementary Materials Sec. II, we show the length scaling behaviors of the YC4 system at different phases.  In Supplementary Materials Sec. III, we show the $T-V$ color figures at different regions of the ground-state phase diagram to support our conclusions of identifying the phases.  In Supplementary Materials Sec. IV, we simulate the order parameter of QAH state using DMRHG, and plot the real-space loop pattern of QAH state. In Supplementary Materials Sec. V, We show the details showing the change of specific heat among the states.  In Supplementary Materials Sec. VI, we show the measurements of purified entanglement entropy, which also helps characterizing the phases. In Supplementary Materials Sec. VII, we show the details of DMRG results for fitting the central charge and measuring the single-particle gap with twisted-boundary conditions in BNDS. In Supplementary Materials Sec. VIII, we plot the energy spectrum from the mean-field hamiltonian with different parameters. In Supplementary Materials Sec. IX, we show the first-order transition between stripe and QAH insulators.
	
	\subsection{Section I: Exponential Tensor Renormalization Group Method and Tangent Space Renormalization Group Method}
	The main idea of exponential tensor renormalization group (XTRG) [8] method is, to first construct
	the initial high-temperature density operator $\hat{\rho_0} \equiv \hat{\rho}(\tau ) = e^{-\tau H} $ with $\tau$ being an exponentially small inverse temperature, which can be obtained with ease via Trotter-Suzuki decomposition [51] or series-expansion methods [52]. Subsequently, we evolve the thermal state exponentially by squaring the density operator iteratively, i.e., $\hat{\rho}_n \cdot \hat{\rho}_n \equiv \hat{\rho}(2^n \tau ) \cdot \hat{\rho}(2^n\tau ) \rightarrow \hat{\rho}_{n+1}$. Following this exponential evolution scheme, one can significantly reduce the imaginary-time evolution as well as truncation steps, and thus can obtain highly accurate low-$T$ data in greatly improved efficiencies.
	
	In XTRG simulations, we measure an observable $\hat{O}$:
	\begin{equation}
		\langle \hat{O} \rangle\ (T)=\frac{\text{Tr}(\hat{\rho}(\frac{\beta}{2})\ \hat{O}\  \hat{\rho}(\frac{\beta}{2}))}{\text{Tr}(\hat{\rho}(\frac{\beta}{2})\hat{\rho}(\frac{\beta}{2}))}
	\end{equation}
	where we write the operator $\hat O$ as an MPO and inverse temperature $\beta=1/T$.
	
	When adapting XTRG to fermion systems, one should take care of the fermionic sign of exchanging two electrons. In this work, we are working on the many-body basis $| n_1\ n_2 \cdot\cdot\cdot n_N \rangle \equiv (c_N^\dagger)^{n_N}\cdot\cdot\cdot(c_2^\dagger)^{n_2}(c_1^\dagger)^{n_1}|\Omega \rangle$, where $n_i \in  \{ 0, 1 \}$ is the number of electrons at the site $i$ and $|\Omega\rangle$ is the vacuum state. Generically in this basis, the one-body operator $c_i^\dagger c_j$ (assuming $j < i$) requires an sign $\prod_{l=j+1}^{i} (-1)^{n_l}$, in addition to transform the state $|n_1 \cdot\cdot\cdot n_j \cdot\cdot\cdot n_i \cdot\cdot\cdot n_N \rangle $ to the state $|n_1 \cdot\cdot\cdot n_{j-1} \cdot\cdot\cdot n_{i+1} \cdot\cdot\cdot n_N \rangle $.
	
	\highlight{
		We also conclude the {\it tan}TRG method [9] here. Using the technique of time-dependent variational principle, for the generation to density operator $\rho$, the optimal tangent vector $X_\rho$ is found on the tangent space $T_\rho \mathcal{M}$, i.e., $\frac{\rho}{\beta}=\mathop{\argmin}_{X_\rho\in T_\rho \mathcal{M}} \|X_\rho+H\rho\|$. For the MPO of $\rho$, the imaginary-time evolution equation can then be expressed as:
		\begin{equation}
			\frac{dA_i}{d\beta}=-H_i^{(1)}A_i+A_i^LH_i^{(0)}S_i,
		\end{equation}
		where $H_i^{(1)}$ is the on-site effective hamiltonian, and $H_i^{(0)}$ is the bond effective hamiltonian. Using splitting method, the equation above can be separated into two linear equations regarding to the site and bond updates: $A_i(\beta+\tau)=e^{-\tau H_i^{(1)}}A_i(\beta)$ and $S_i(\beta+\tau)=e^{\tau H_i^{(0)}}S_i(\beta)$. Through site-wise sweeping, the integrator guarantees an optimal approximation within the MPO manifold $\mathcal{M}$.
	}
	
	\subsection{Section II: Length Scaling of The YC4 System}
	In the main text, most of the data are from simulation of YC$4\times 16$ cylinders. Here, we show the length scaling behaviors.  As shown in Fig. \ref{length_scaling}, we simulate $L=12,\ 16,\ 20$ for QAH structure factor, bond nematic order, site nematic order and stripe structure factor respectively, which signify the length we take in the main text is sufficient for analysis. 
	\begin{figure}[H]
		\centering
		\includegraphics[width=0.98\textwidth]{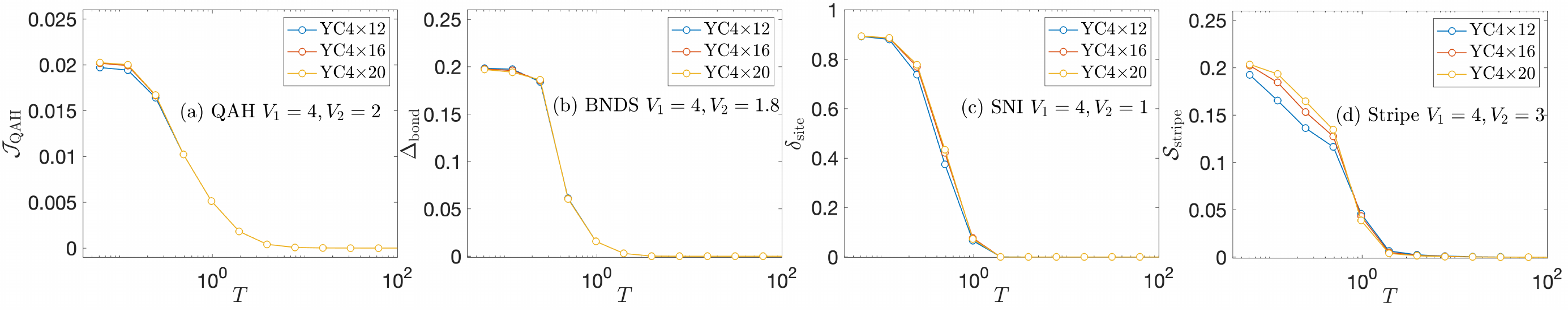}
		\caption{For each of the phase, the length scaling behavior is shown using their order parameters or structure factors, including (a) QAH, (b) BNDS, (c) SNI, and (d) Stripe respectively.}
		\label{length_scaling}
	\end{figure}

	\subsection{Section III: Thermodynamic results of other cuts  on the $V_1-V_2$ phase diagram }
	Firstly, we show the thermodynamic measurements in the setting of the previous work on bond order phase in this model~[5], without NNN interaction, in Fig.~\ref{colorfig_v2_0}. The results further support the previous conclusion that there exists a bond-orders phase between QAH and SNI.
	\begin{figure}[H]
		\centering
		\includegraphics[width=\textwidth]{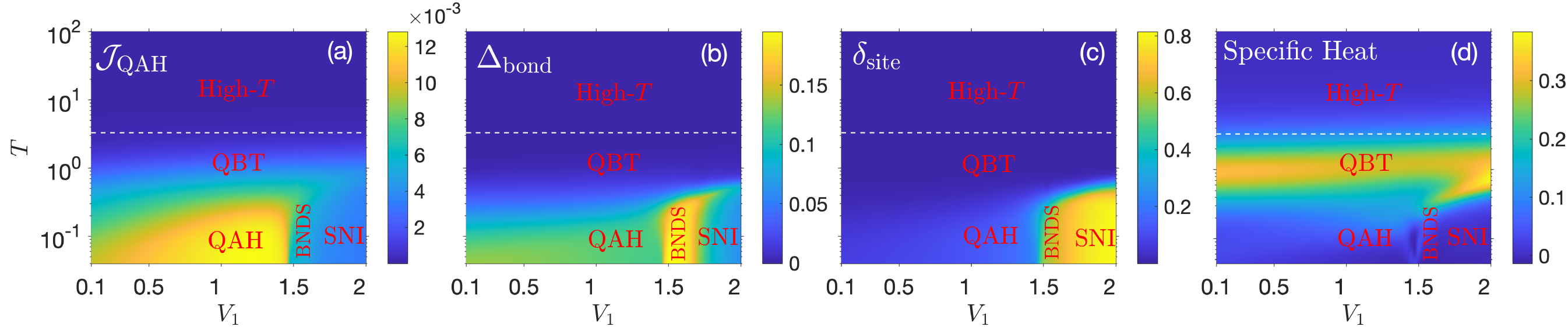}
		\caption{With fixed $V_2=0$ and changing $V_1$, the QAH structure factor, bond order, site nematic order, and specific heat are plotted in (a)-(d), whose x-axis is $V_2$ and y-axis is $T$.}
		\label{colorfig_v2_0}
	\end{figure}
	In the main text, we have shown that, the bond-ordered dirac semimetal exists not only in the NN interaction region, but also in the interplay of NN and NNN interactions. We add more data to support our conclusion and ground-state phase diagram by showing thermodynamic measurements of other cuts on the $V_1-V_2$ parameter space, such as fixing $V_1=2$ and change $V_2$, which is shown in Fig.\ref{colorfig_v1_2}.
	\begin{figure}[H]
		\centering
		
		\includegraphics[width=\textwidth]{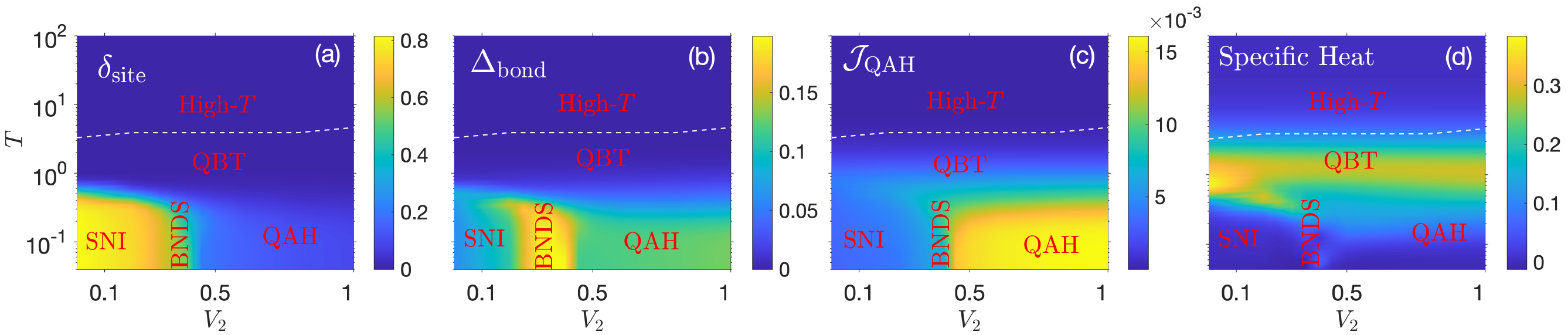}

		\caption{With fixed $V_1=2$ and changing $V_2$ (not including the Stripe region),the QAH structure factor, bond order, site nematic order, and specific heat are plotted in (a)-(d), whose x-axis is $V_1$ and y-axis is $T$. }
		\label{colorfig_v1_2}
	\end{figure}
	
	From the ground-state phase diagram, we can see the QAH region is getting narrow in large $V$, so we also show the thermodynamic simulations of fixed $V_1=5$ and changing $V_2$ in Fig.\ref{colorfig_v1_5}
	\begin{figure}[H]
		\centering
		
		\includegraphics[width=\textwidth]{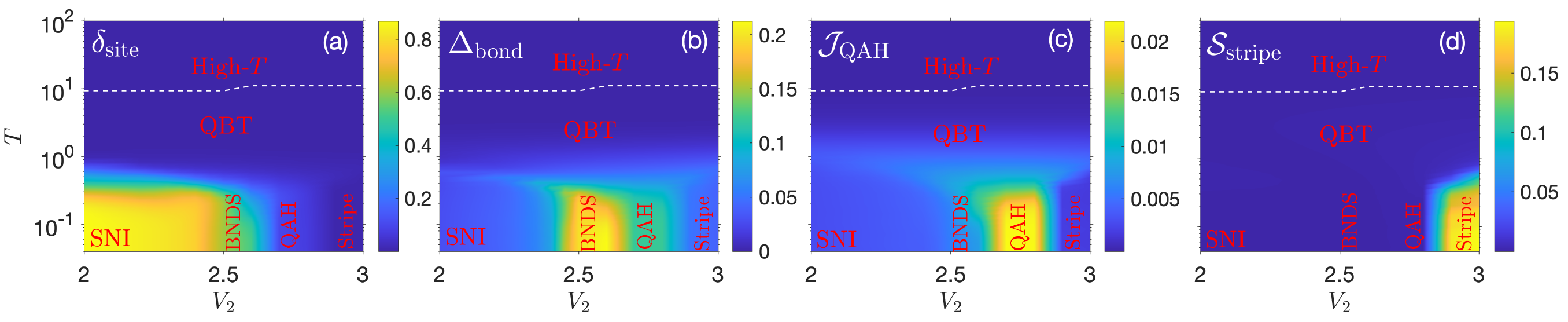}

		\caption{With fixed $V_1=5$ and changing $V_2$, the QAH structure factor, bond order, site nematic order, and stripe structure factor are obtained in (a)-(d), whose x-axis is $V_2$ and y-axis is $T$.}
		\label{colorfig_v1_5}
	\end{figure}
	
	\subsection{Section IV: Crossover between the semimetal and high-$T$ regime }
	\begin{figure}[H]
		\centering
		\includegraphics[width=\textwidth]{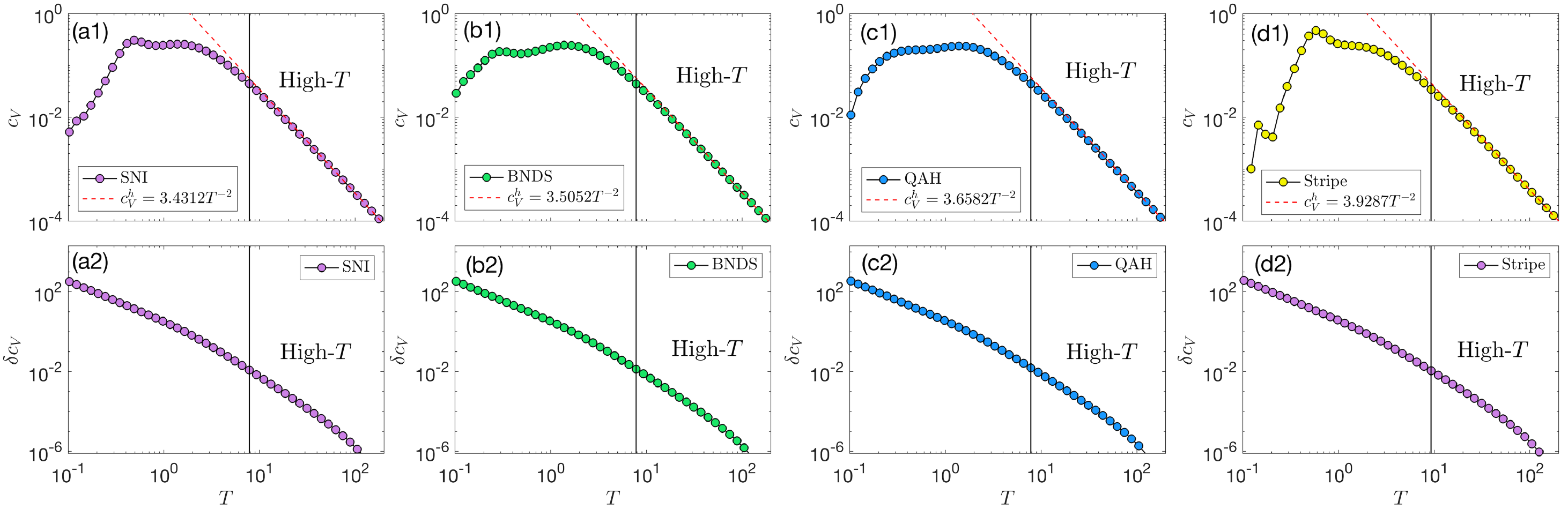}
		\caption{In YC$4\times12$ system with fixed $V_1=4$ for (a1) SNI at $V_2=1.65$, (b1) BNDS at $V_2=1.85$, (c1) QAH at $V_2=2.2$, and (d1) Stripe at $V_2=2.7$, specific heat $c_V$ curves are shown versus $T$ in a double-logarithmic scale. At high-temperature limit, $c_V$ behaves as $\sim T^{-2}$, and is fitted by the red dashed line $c_V^h$. Lower panels show the error defined as $\delta c_V=|c_V-c_V^h|$ as function of $T$. The crossover temperature between high-$T$ and intermediate semimetal is determined by a threshold $\epsilon =10^{-2}$. }
		\label{cv_error}
	\end{figure}
	In this section, we discuss the crossover behavior between intermediate semimetal regime and high-$T$ gas-like regime. In a high $T=1/\tau$(i.e., a small inverse temperature), the internal energy density of the system is 
	\begin{equation}
		u(\tau)=\frac{1}{N}\frac{\text{Tr}[H\cdot \rho(\tau)]}{\text{Tr}[\rho(\tau)]}=\frac{1}{N}\frac{\text{Tr}(H\cdot e^{-\tau H})}{\text{Tr}(e^{-\tau H})},
	\end{equation}
	via Taylor expansion, can be expressed as 
	\begin{equation}
		u(\tau)=\frac{1}{N}\left[\frac{\text{Tr}(H)}{Z^0}-\frac{\text{Tr}(H^2)}{Z^0}\tau+(\frac{\text{Tr}(H)}{Z^0})^2\tau+O(\tau^2)\right].
	\end{equation}
	Thus in the large-$T$ limit, it yields
	\begin{equation}
		c_V\equiv\partial u/\partial T=(\partial u/\partial \tau)\cdot(\partial \tau/\partial \tau/\partial T)\sim T^{-2}
	\end{equation}
	for specific heat.
	As shown in Fig.\ref{cv_error}, we show the specific heat of the system at 4 different phases, which exhibit predominant power-law decay at high-$T$ regime. We fit the high-$T$ regime by $c_V^h$ respectively and compute the error between them. We classify the high-$T$ regime where the error $\delta c_V < \epsilon=10^{-2}$.

	\subsection{Section V: Order parameter and NN loop current of QAH state}
	For more evidence of QAH state in our phase diagram, we show the DMRG measurement of the order parameter $\Delta_{\mathrm{QAH}}=-4i(c_i^\dagger c_j^{\ } -c_j^\dagger c_i^{\ } )$ with fixed $V_1=4$ as shown in Fig.\ref{qah_order}, which agrees with the finite-temperature phase diagram in the main text. 
	\begin{figure}[H]
		\centering
		\includegraphics[width=0.5\textwidth]{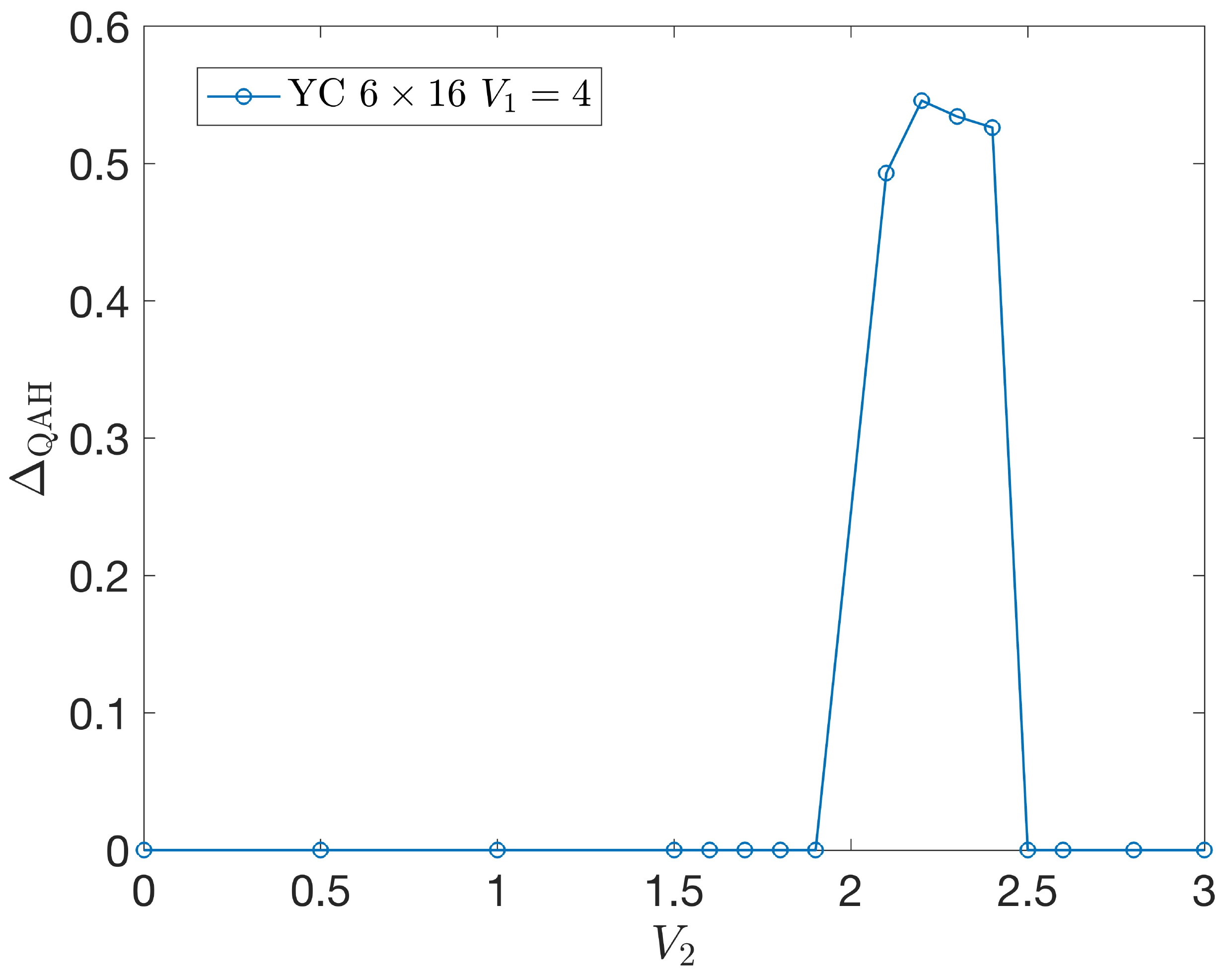}
		\caption{YC6 DMRG simulations of QAH order parameter with fixed $V_1=4$.}
		\label{qah_order}
	\end{figure}
	In this section, we also show the real-space pattern of loop current in QAH state as we have measured the time-reversal symmetry breaking order parameter in the main text.
	
	\begin{figure}[H]
		\centering
		\includegraphics[width=\textwidth]{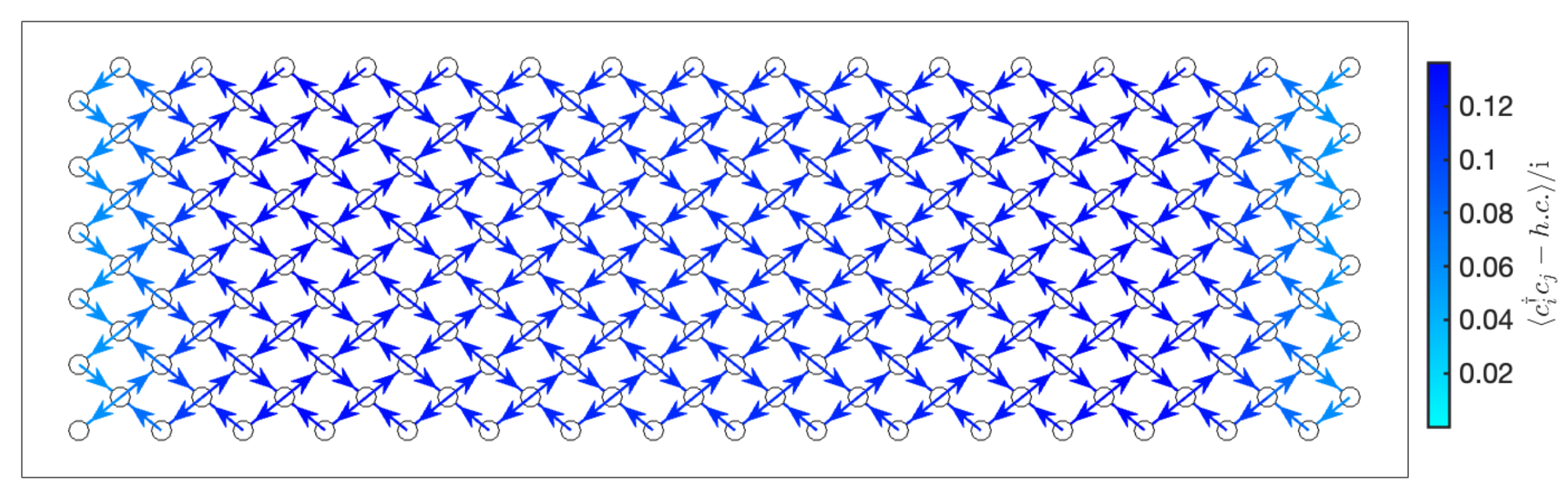}
		\caption{The NN loop pattern of YC $6\times 16$ cylinder with $V_1=4$ and $V_2=2.2$.}
		\label{loop_qah}
	\end{figure}
	
	The exact loop current is plotted in Fig.\ref{loop_qah}, which further clearly supports the evidence of time-reversal symmetry breaking.

	\subsection{Section VI: Purified entanglement entropy }
	The entanglement entropy $S_E$ is measured from the partial trace of the ``superdensity" matrix $\mathcal{R}$ of the ``superstate" $|\Psi|\rangle\equiv \frac{1}{\sqrt{\mathcal{Z}(\beta)}} |e^{-(\beta/2)H}\rangle$, which vectorizes the MPO for $|e^{-(\beta/2)H}\rangle$.
	\begin{figure}[H]
		\subfigure[]{
			\includegraphics[width=0.24\textwidth]{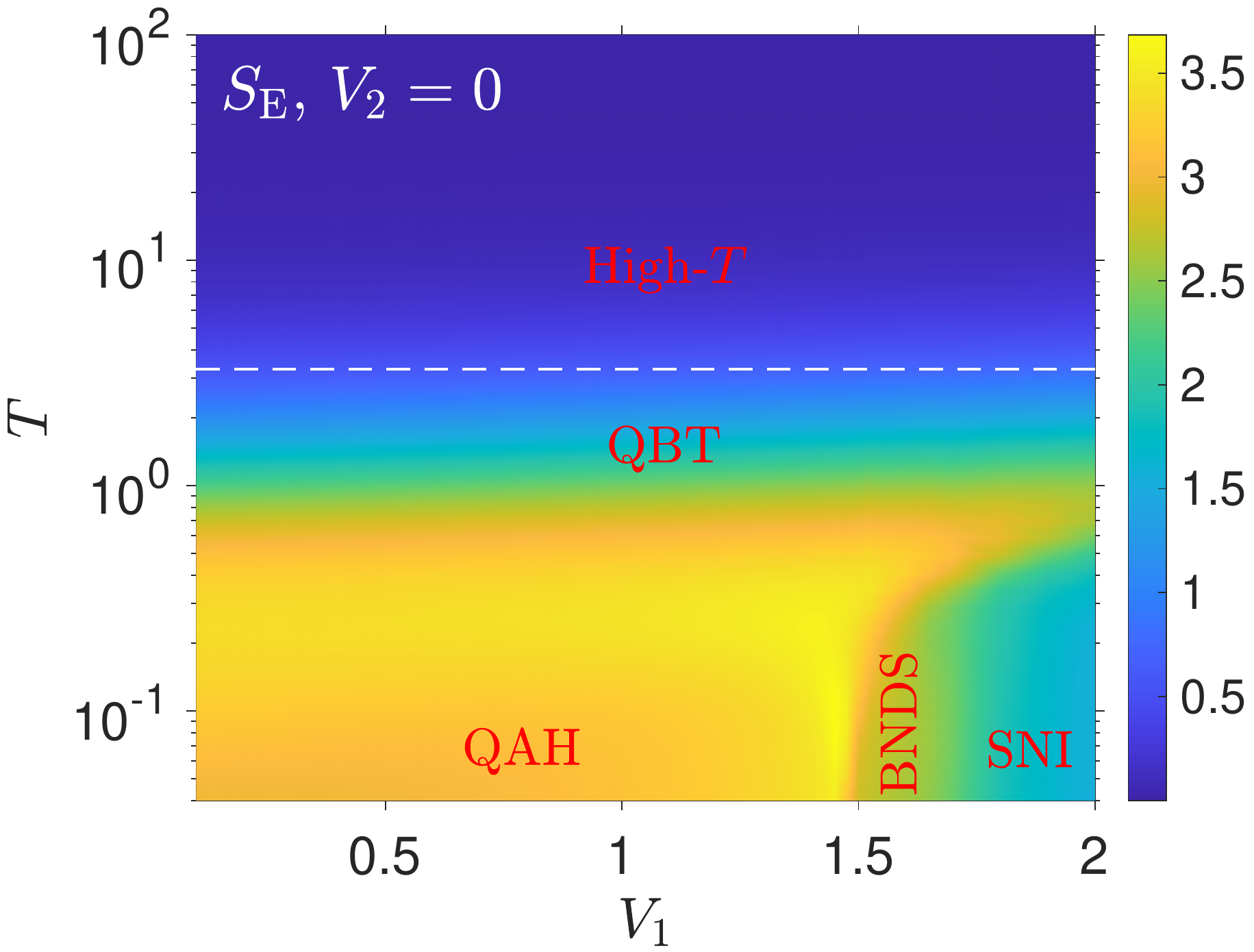}}
		\subfigure[]{
			\includegraphics[width=0.24\textwidth]{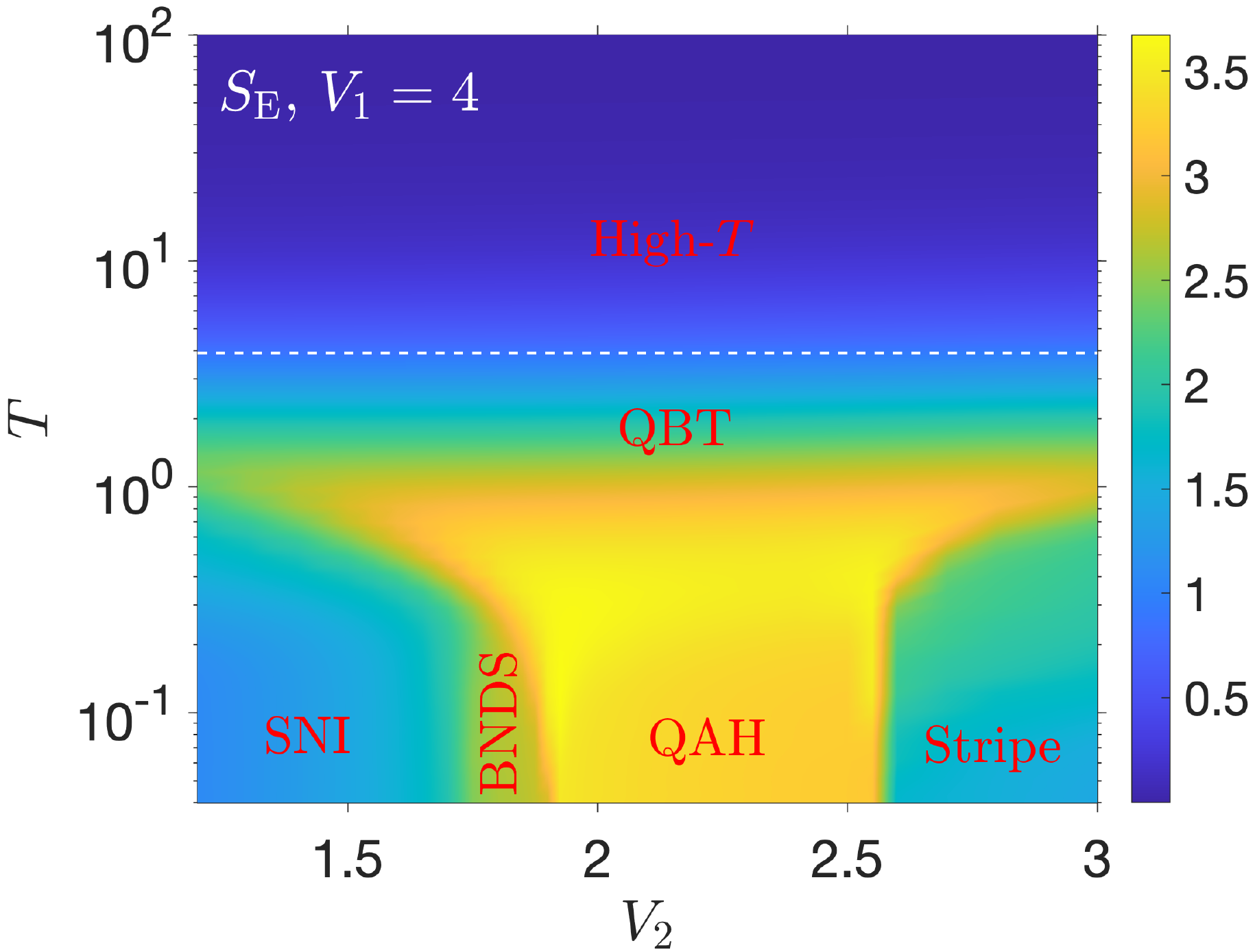}}
		\subfigure[]{	
			\includegraphics[width=0.24\textwidth]{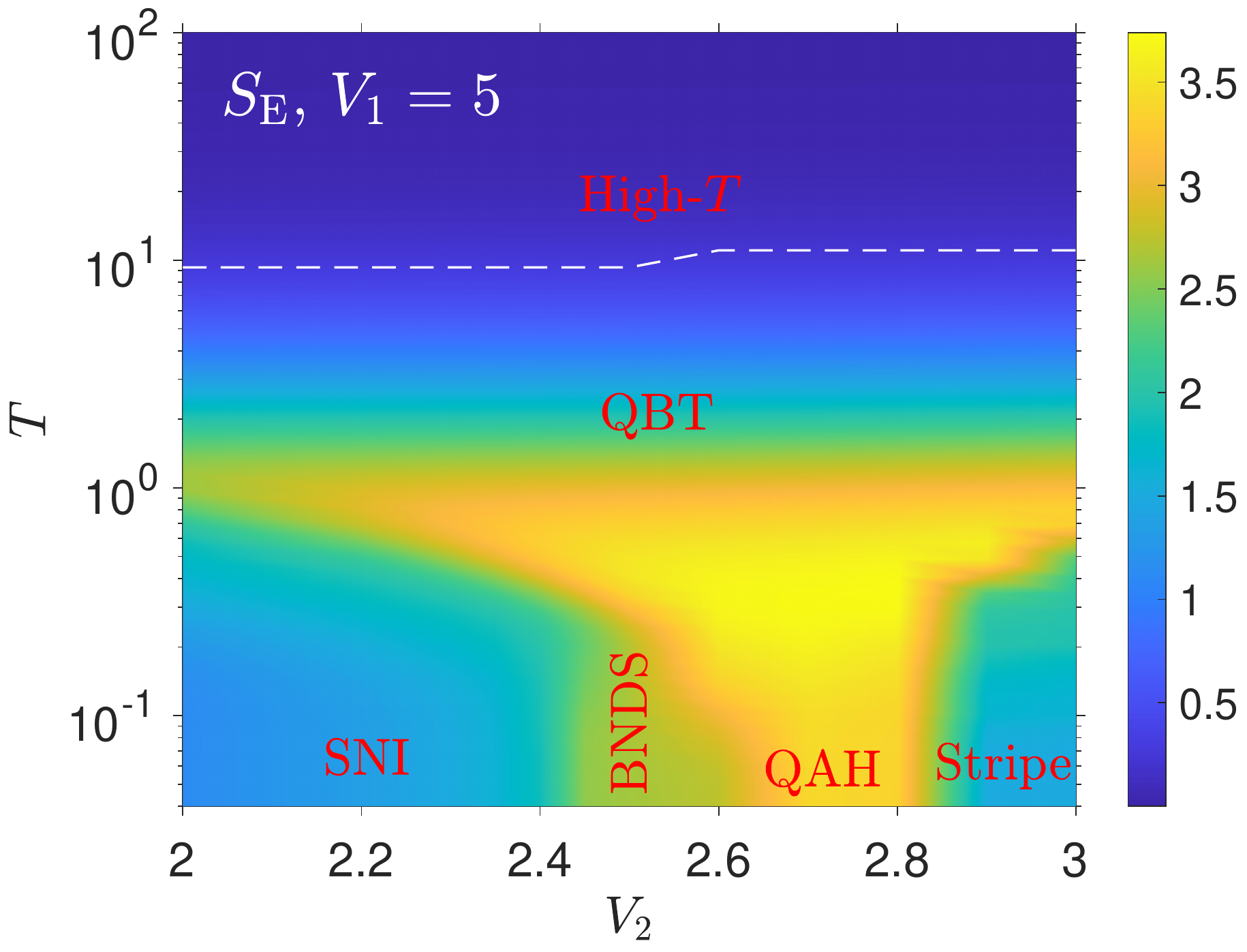}}
		\subfigure[]{
			\includegraphics[width=0.24\textwidth]{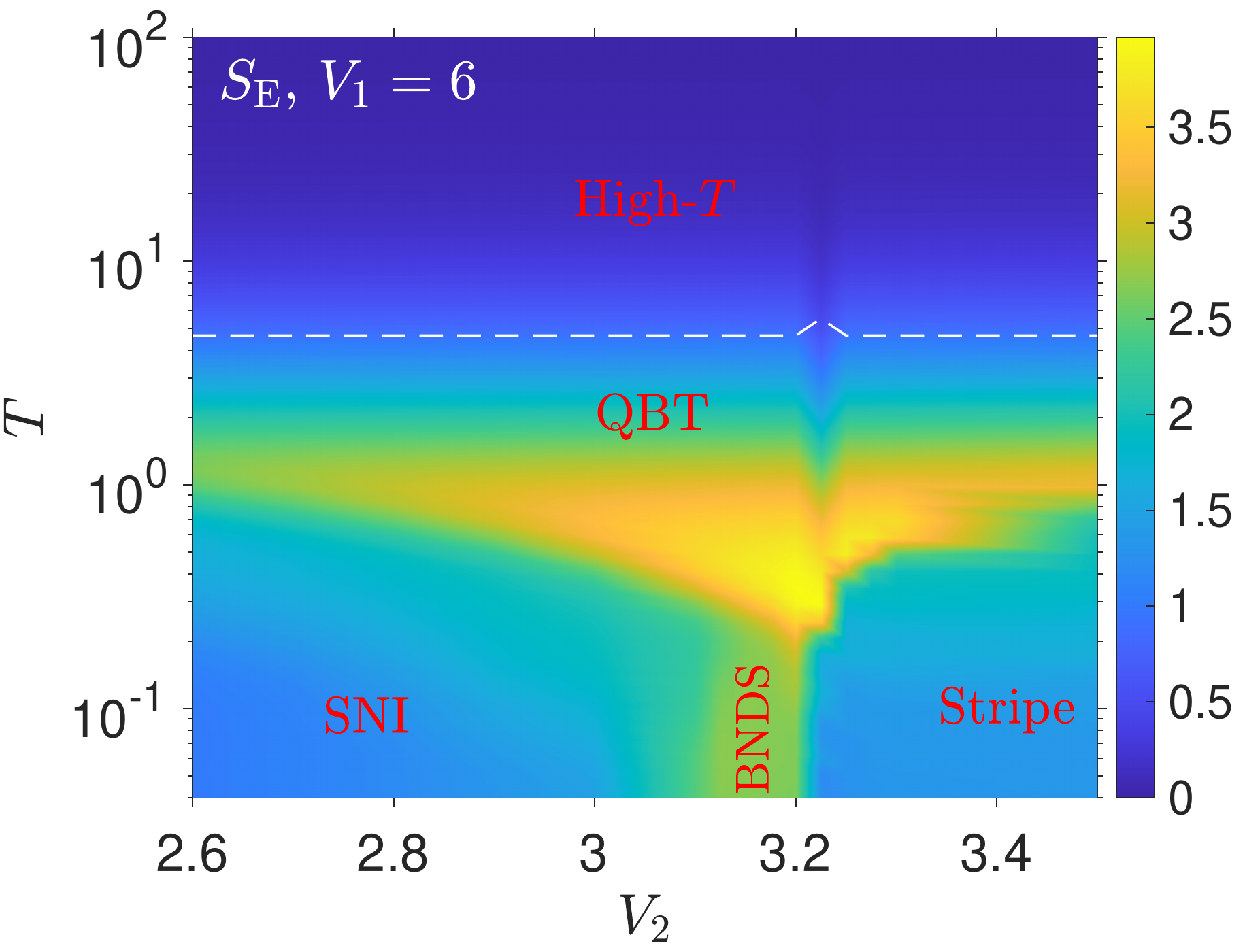}}	
		\caption{The purified entanglement entropy of different $T-V$ regions.}
		\label{purified_se}
	\end{figure}
	
	As shown in Fig.\ref{purified_se}, at low $T$, the entanglement entropy is the highest for QAH (which might be due to the edge states), while much lower for the site-nematic insulator and stripe insulator. The bond-nematic dirac semimetal phase shows higher entanglement entropy than the site nematic insulator, and show clear phase boundaries at low temperature. Such entanglement data indicate a possible gapless state, which are in agreement with previous conclusions.

	\subsection{Section VII: DMRG results in BNDS }
	As the BNDS between stripe insulator and SNI has not been investigated before, we implement DMRG simulations to varify the dirac semimetal phase. Since the dirac points are not on the high-symmetry points in momentum space, we take the cylinder width $W=6$ to decrease the finite-size effect. We simulate the entanglment entropy in the BNDS region and fit it to the universal scaling function $S_{\mathrm{E}}(x)=\frac{c}{6}\log(\frac{L}{\pi}\sin{\frac{x\pi}{L}})$. We show the results in Fig.\ref*{se_bnds}.  With up to $D=4096$, the entanglement entropy is still not converged, with the fitted central charge as large as $c\approx 1.75$, which is shown in the main text and this is enough to support the gapless BNDS phase.
	
	As shown in the main text, we have measured the single-particle gap with twisted-boundary conditions in BNDS phase with $V_1=6$ and $V_2=3.185$. We define the gap $\Delta_{\mathrm{sp}}=E_e-E_g$, where $E_g$ is the ground state energy and $E_e$ is the lowest energy in the single-particle-excited sector of our DMRG simulations. We do extrapolations with different bond dimensions to obtain the energies as shown in Fig.\ref{extrap_e};
	\begin{figure}[H]
		\centering
		\subfigure[YC6$\times 12$]{
			\includegraphics[width=0.4\textwidth]{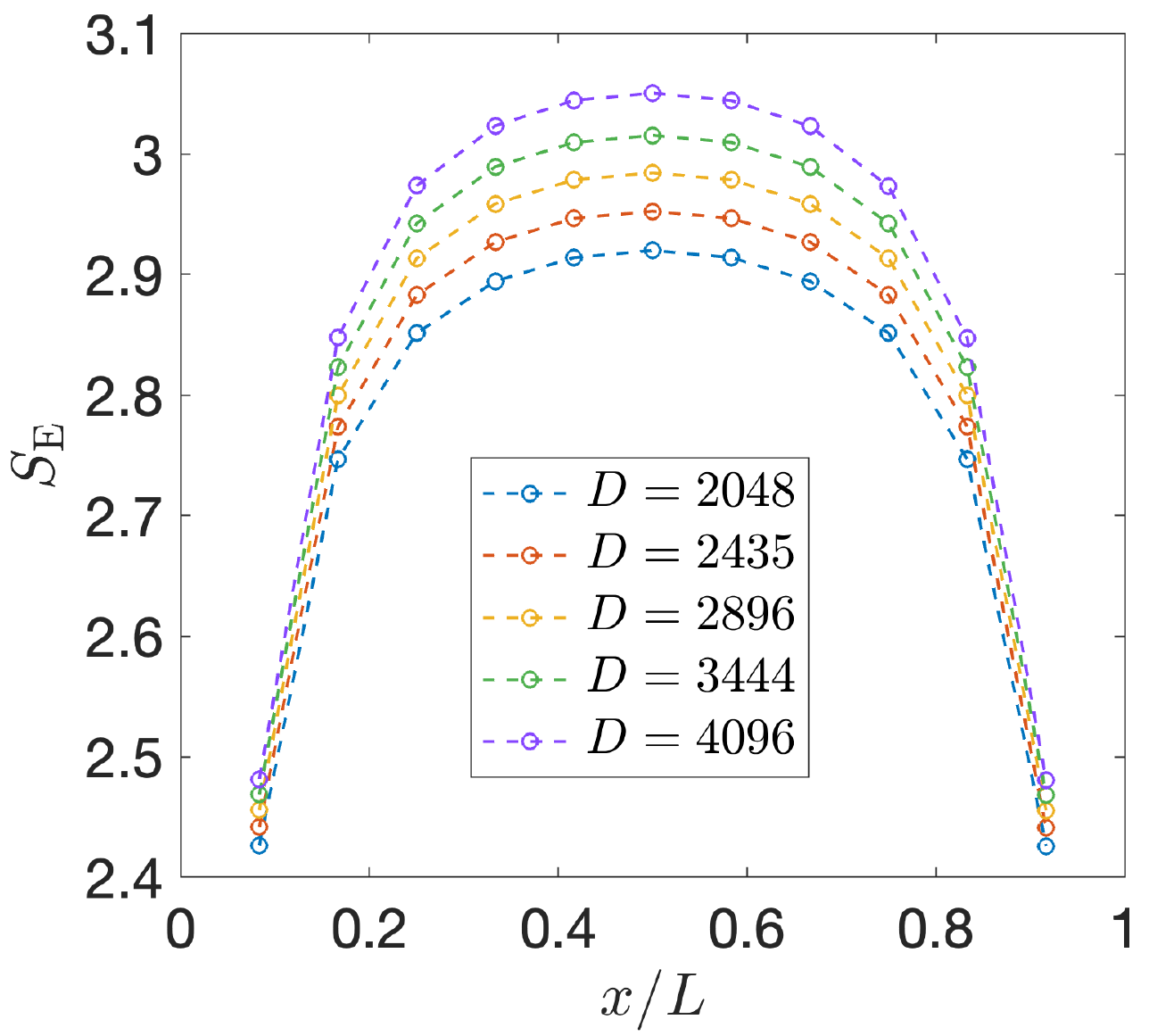}}
		\subfigure[YC6$\times 14$]{
			\includegraphics[width=0.4\textwidth]{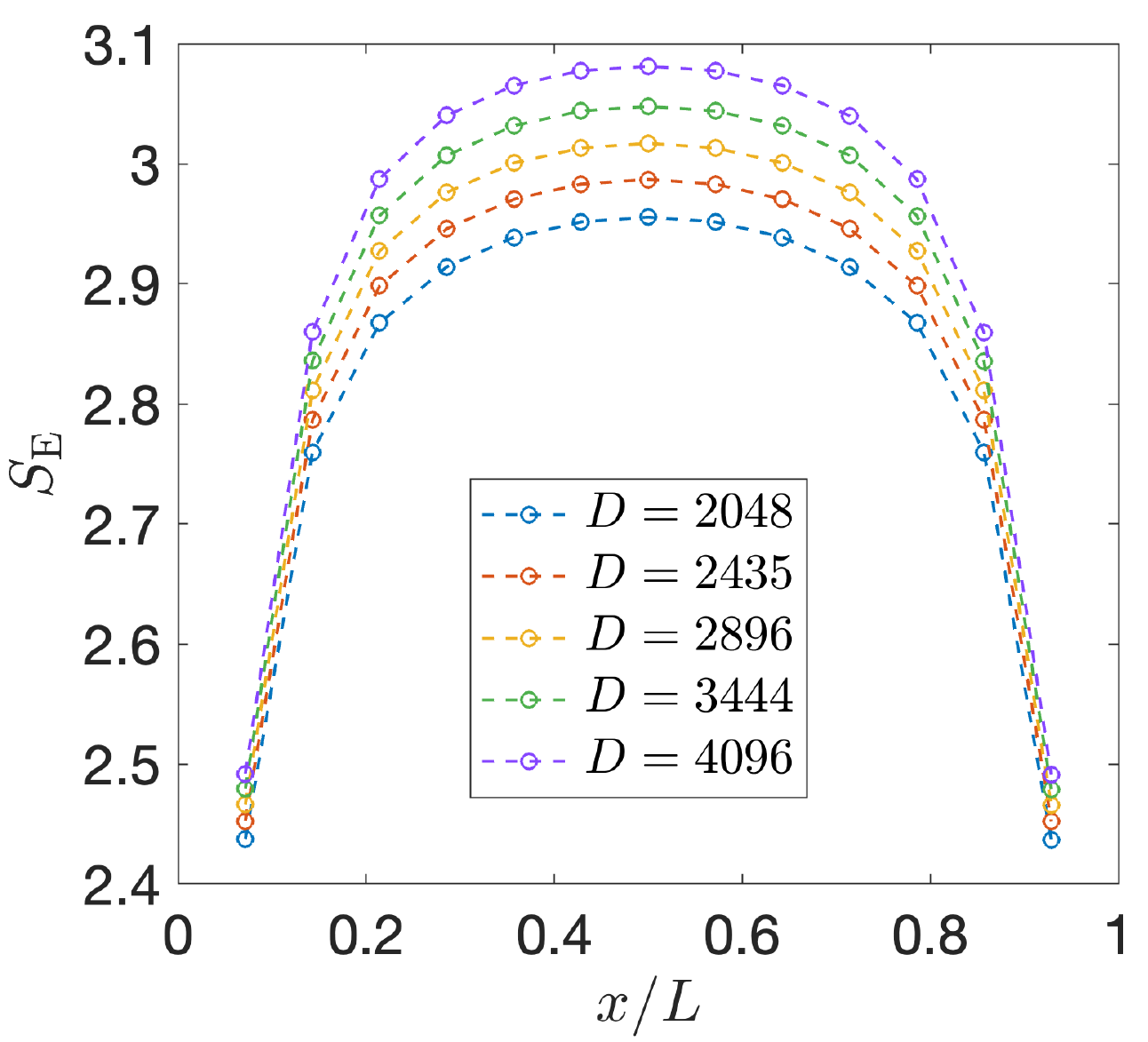}}
		
		\caption{The entanglement entropy obtained by DMRG in the BNDS region with $V_1=6$ and $V_2=3.185$ with different bond dimensions.  }
		\label{se_bnds}
	\end{figure}
	
	\begin{figure}[H]
		\centering
		\subfigure[]{
			\includegraphics[width=0.4\textwidth]{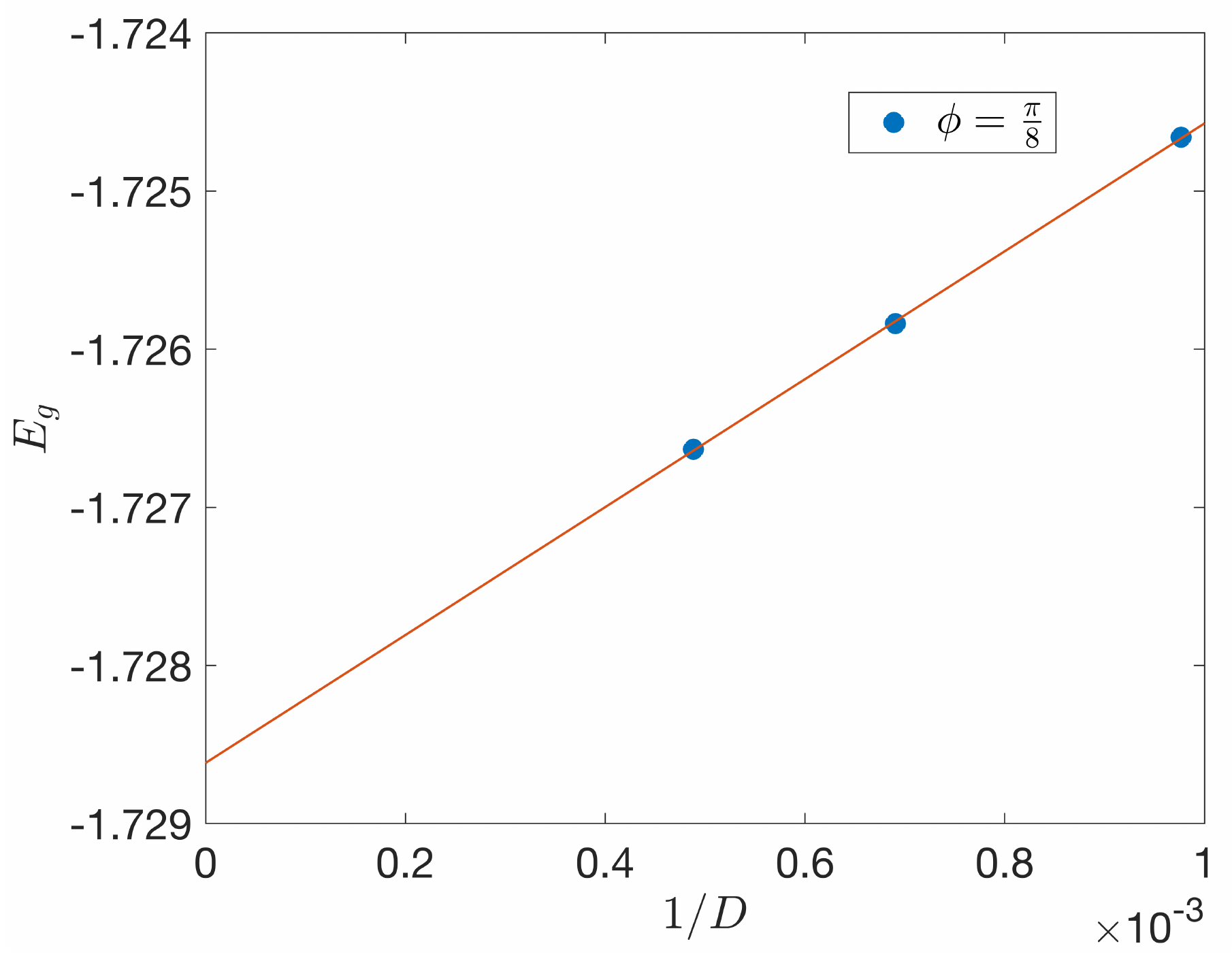}}
		\subfigure[]{
			\includegraphics[width=0.4\textwidth]{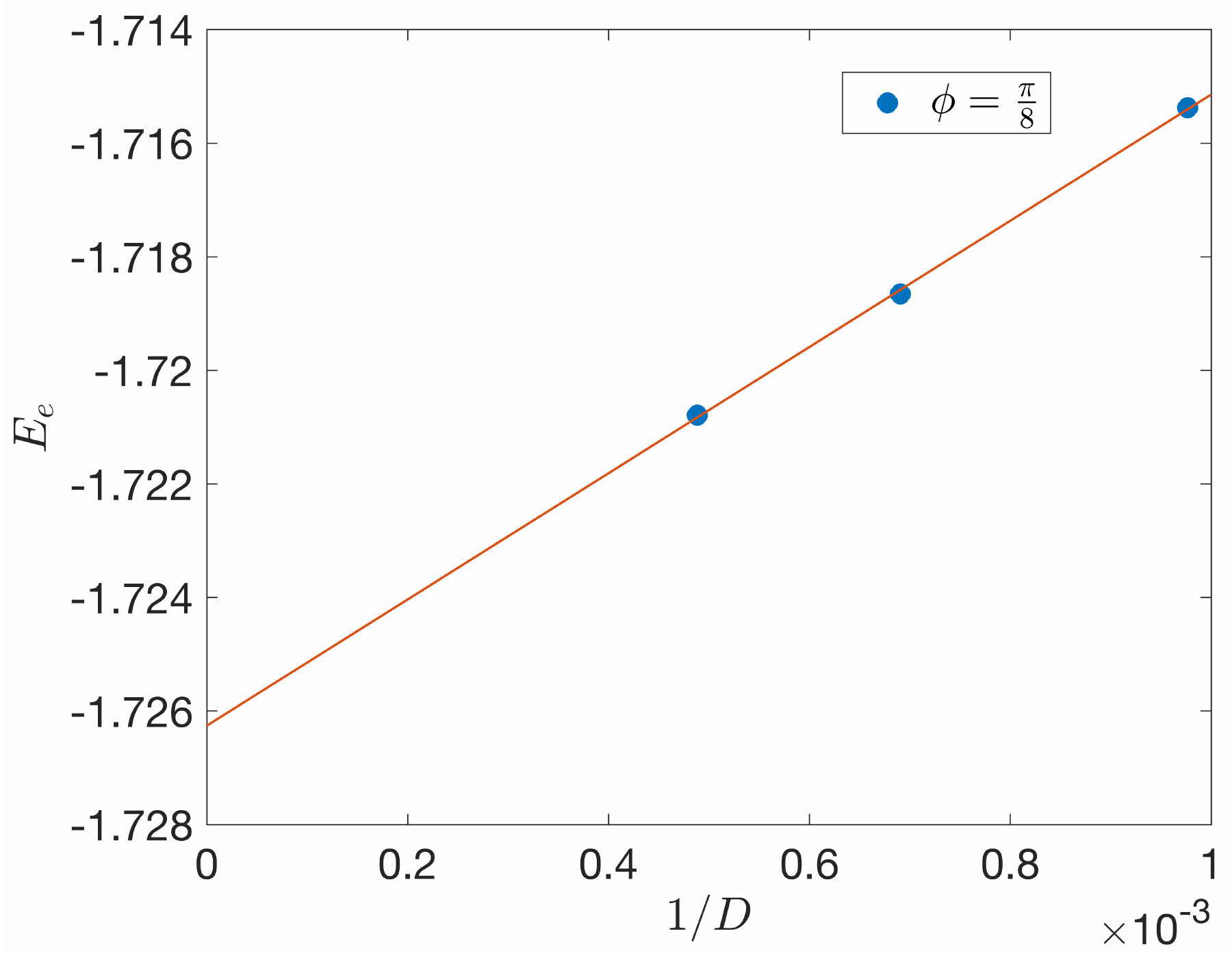}}
		
		\caption{The extrapolated ground-state energy (a) and single-particle-exited energy (b) with $V_1=6$, $V_2=3.185$, and flux $\phi=\pi/8$.  }
		\label{extrap_e}
	\end{figure}
	
	\subsection{Section VIII: Energy spectrum of mean field hamiltonian }
	For the mean field analysis, we write the hamiltonian as: 
	\begin{equation}
		\mathcal{H}(k)=\begin{gathered}
			\begin{pmatrix}
				c_a^\dagger(k) & c_b^\dagger(k)
			\end{pmatrix}
			\begin{pmatrix}
				(t'+\frac{\Delta}{2})\cos(k_x)+(-t'+\frac{\Delta}{2})\cos(k_y)-\frac{\delta}{2}& t\cdot \cos(\frac{k_x+k_y}{2})e^{-i\phi}+t\cdot \cos(\frac{k_x-k_y}{2})e^{i\phi}
				\\   t\cdot \cos(\frac{k_x+k_y}{2})e^{i\phi}+t\cdot \cos(\frac{k_x-k_y}{2})e^{-i\phi}&(-t'-\frac{\Delta}{2})\cos(k_x)+(t'-\frac{\Delta}{2})\cos(k_y)+\frac{\delta}{2}
			\end{pmatrix}
			\begin{pmatrix}
				c_a(k) \\ c_b(k)
			\end{pmatrix}.
		\end{gathered}
		\label{tightbinding}
	\end{equation}
	From the tight-binding part of Eq.~(1), 
	we change NN hopping $t\to t e^{\pm i\phi}$ with the sign structure $\{\pm\phi\}$ 
	following the direction shown in Fig.~1(c) to mimic the QAH phase;
	For the bond order state, we change NNN hopping $t'\to (t'-\epsilon_i\Delta)$ with positive $\Delta$, 
	i.e. stronger hopping integral along vertical $\mathbf{a_2}$ direction; 
	For the site-nematic state, we add sublattice chemical potential term 
	$\delta \sum_\mathbf{r}(n_{\mathbf{r},A} - n_{\mathbf{r},B})$
	to the hamiltonian Eq.~1.
	we plot the energy spectrum as shown in Fig.\ref{energy_band}. For the original tight-binding model, there is a QBT at ($\pi,\pi$). In the QAH with finite NN loop current, the gap will open immediately. Besides, we estimate the mean-field gap in QAH through this effective hamiltonian. For example, when $V_1=6$ and $V_2=3.185$, we have measured in our DMRG simulations that the ratio of NN bond current and NN hopping amplitude (measured to be around 0.18) is approximately 0.35, so we tune $\phi=0.3367$ here and the gap is around 0.22, as shown in Fig.\ref{energy_band}(b). When adding site nematic order or NNN bond nematic order into the hamiltonian, the QBT will split into two dirac touching points, and the larger the order, the farther are the two points~[1,6]. If the two dirac points getting far enough (qualitatively in this model, $\delta > 2$), the system will be gapped, while the bond nematic order itself cannot open the gap for this system and even when the bond-nematic order is almost the same as the NNN hopping amplitude in the hamiltonian, the moving distance of Dirac cones is less than $\pi/2$. 
	\begin{figure}[H]
		\centering
		\subfigure[$\phi=0, \delta=0, \Delta=0$]{
			\includegraphics[width=0.192\textwidth]{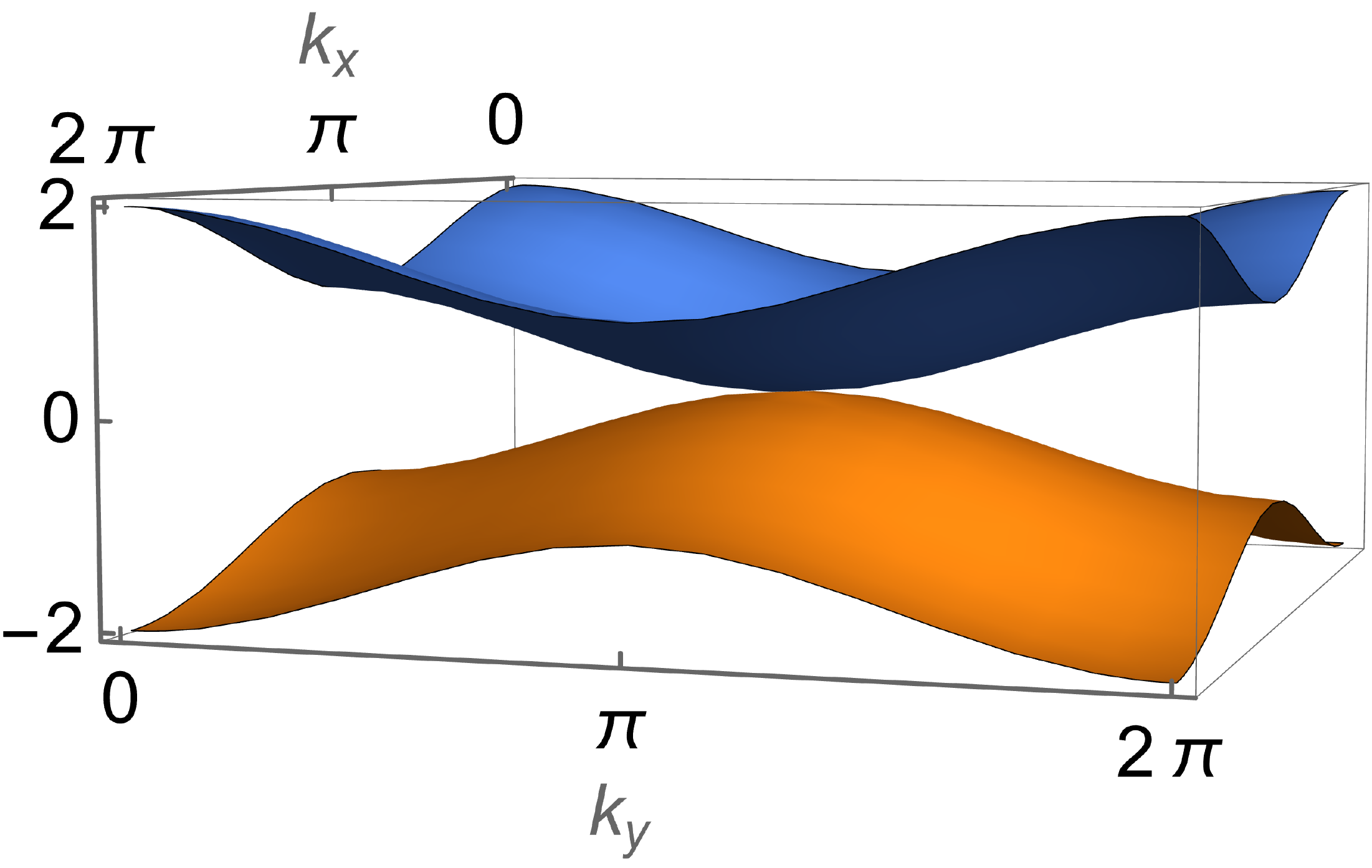}}
		\subfigure[$\phi=0.3367, \delta=0, \Delta=0$]{
			\includegraphics[width=0.192\textwidth]{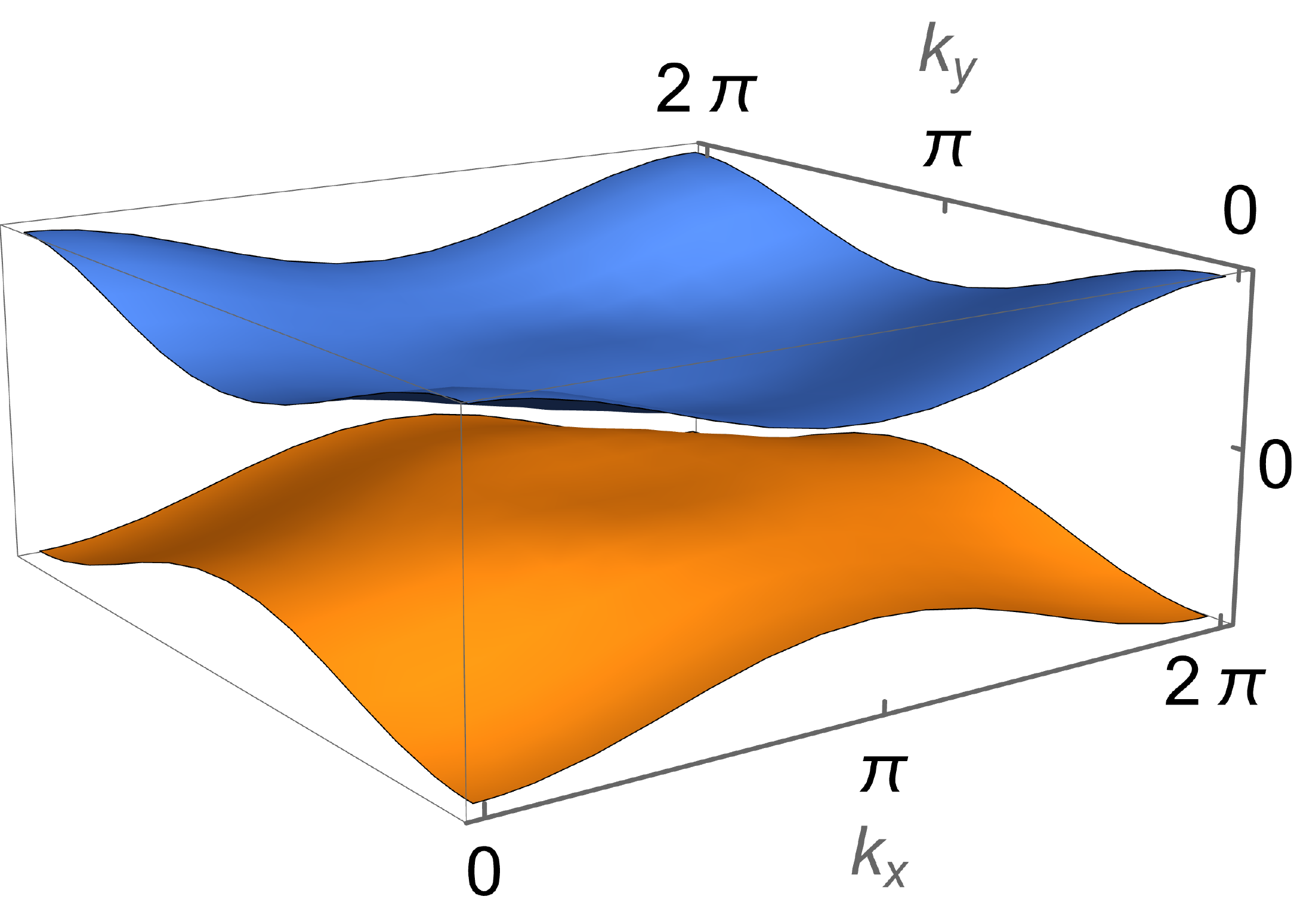}}
		\subfigure[$\phi=0, \delta=0, \Delta=0.9$]{
			\includegraphics[width=0.182\textwidth]{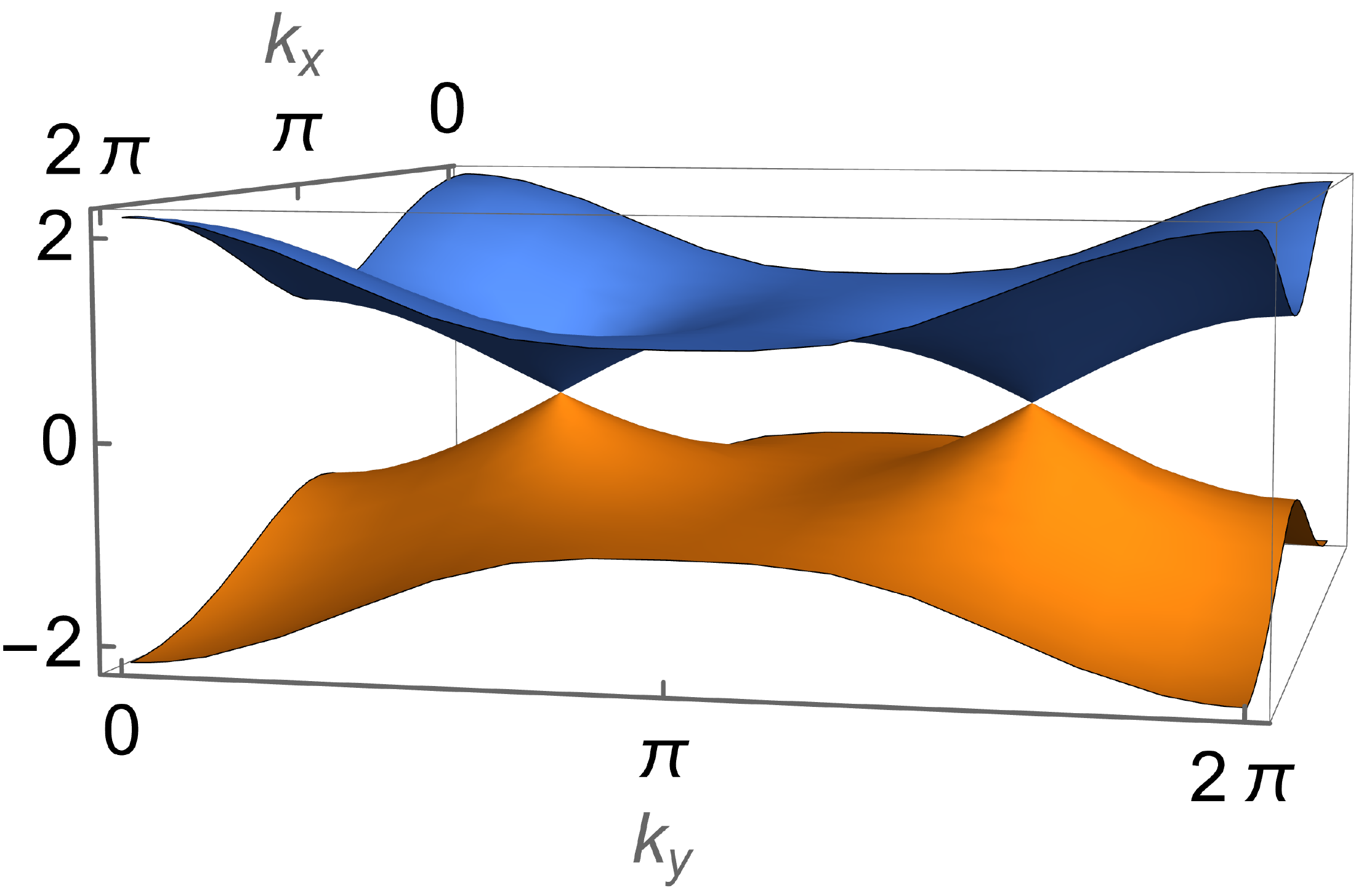}}
		\subfigure[$\phi=0, \delta=0.9, \Delta=0$]{
			\includegraphics[width=0.182\textwidth]{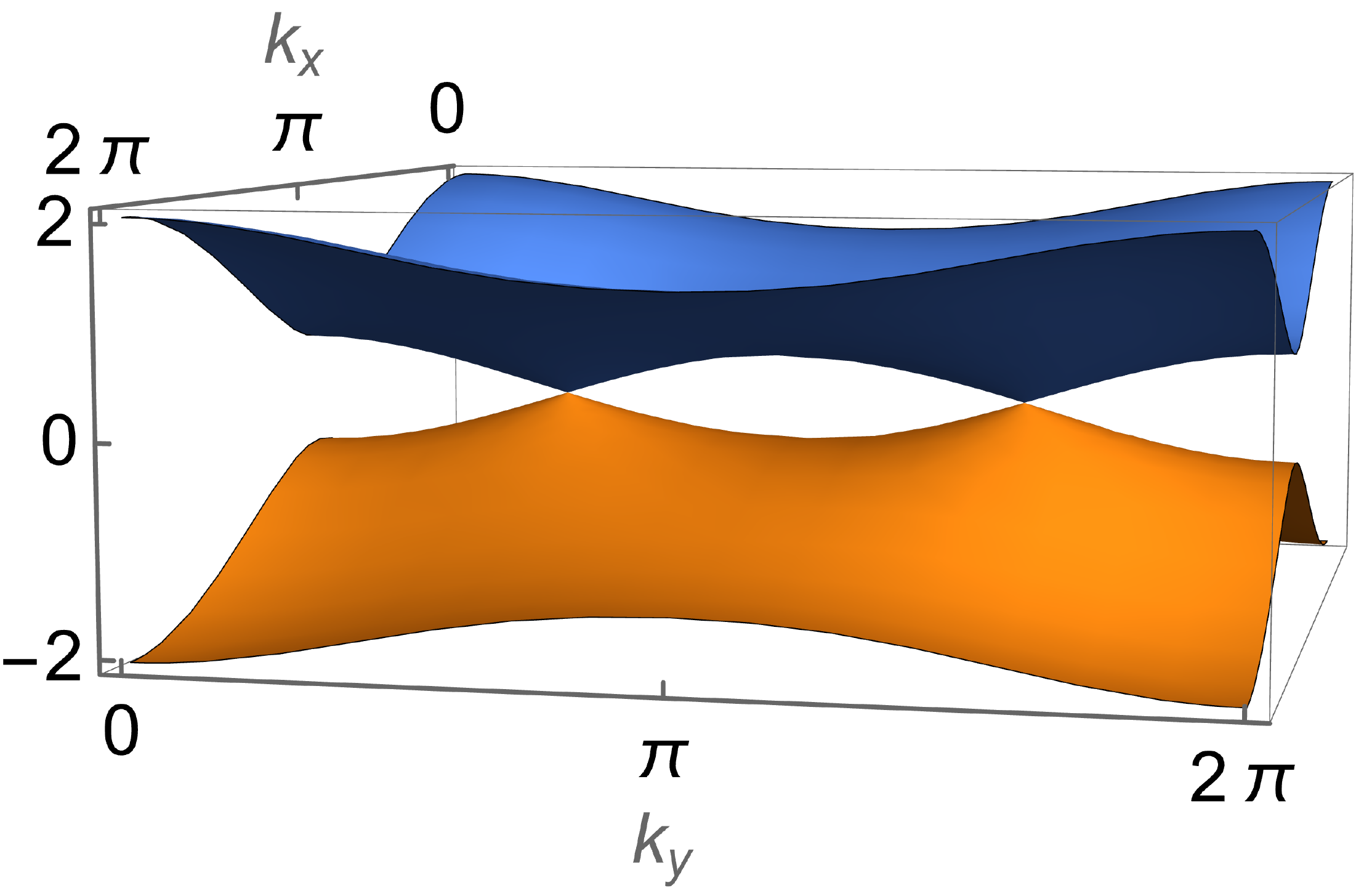}}
		\subfigure[$\phi=0, \delta=2.5, \Delta=0$]{
			\includegraphics[width=0.182\textwidth]{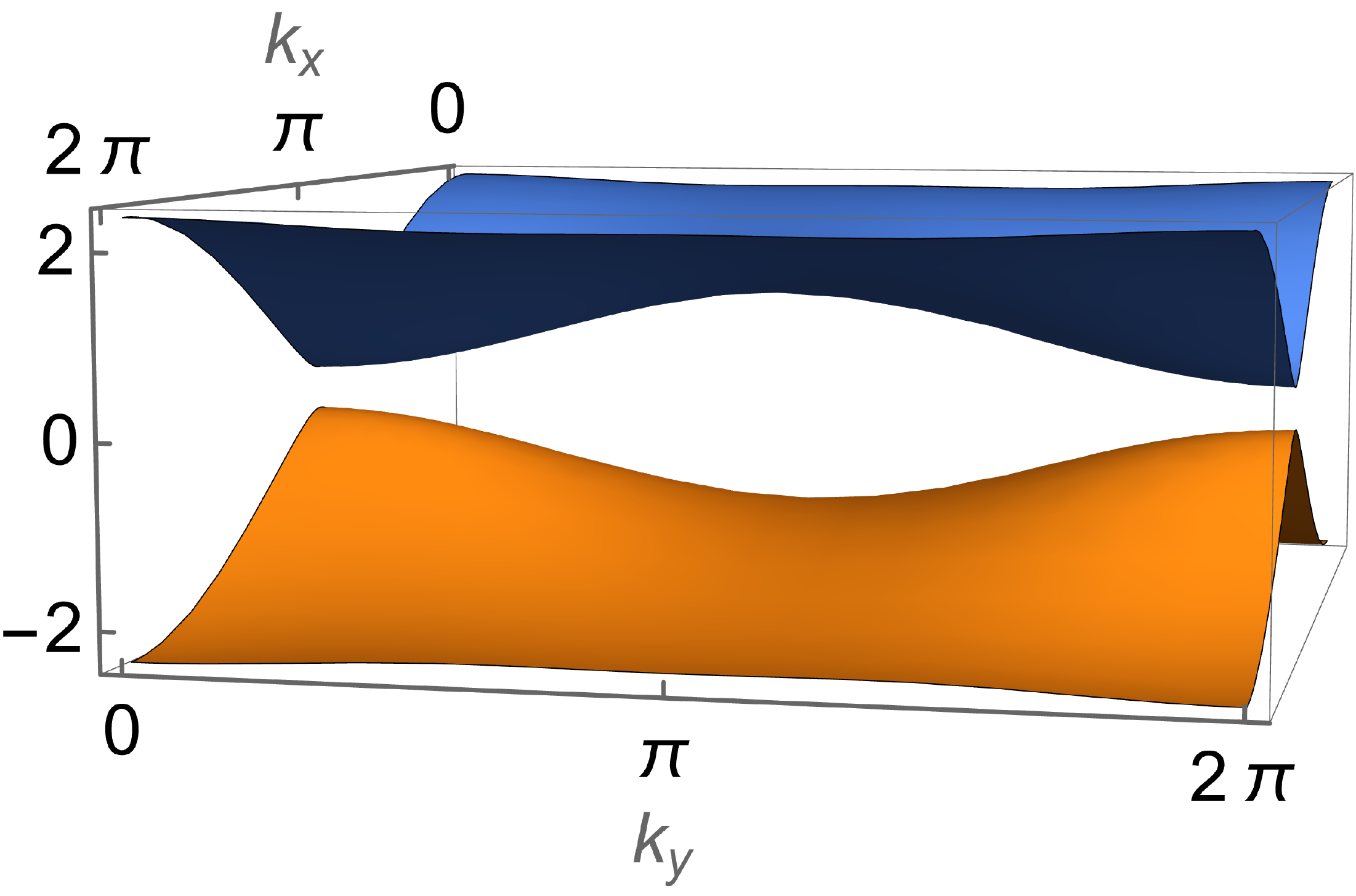}}
		
		\caption{The energy spectrum of the mean field hamiltonian. (a) QBT. (b) QAH with NN loop current. (c) Bond nematic semimetal. (d) Site nematic semimetal. (e) Site nematic insulator. }
		\label{energy_band}
	\end{figure}
	
	As we have discussed in the main text, the NNN bond nematic order and site nematic order break the same symmetry, we show how they can induce each other in the mean-field level as shown in Fig.\ref{mf_nematic_orders}. Besides, when $\delta$ is large enough and the the system is gapped out, the bond nematic order $\Delta$ will drop. When $\delta=2$ in the mean-field hamiltonian, the two Dirac points merge into each other and are about to gap out. At this point, the bond order drop most significantly.
	
	\begin{figure}[H]
		\centering
		
		\includegraphics[width=0.7\textwidth]{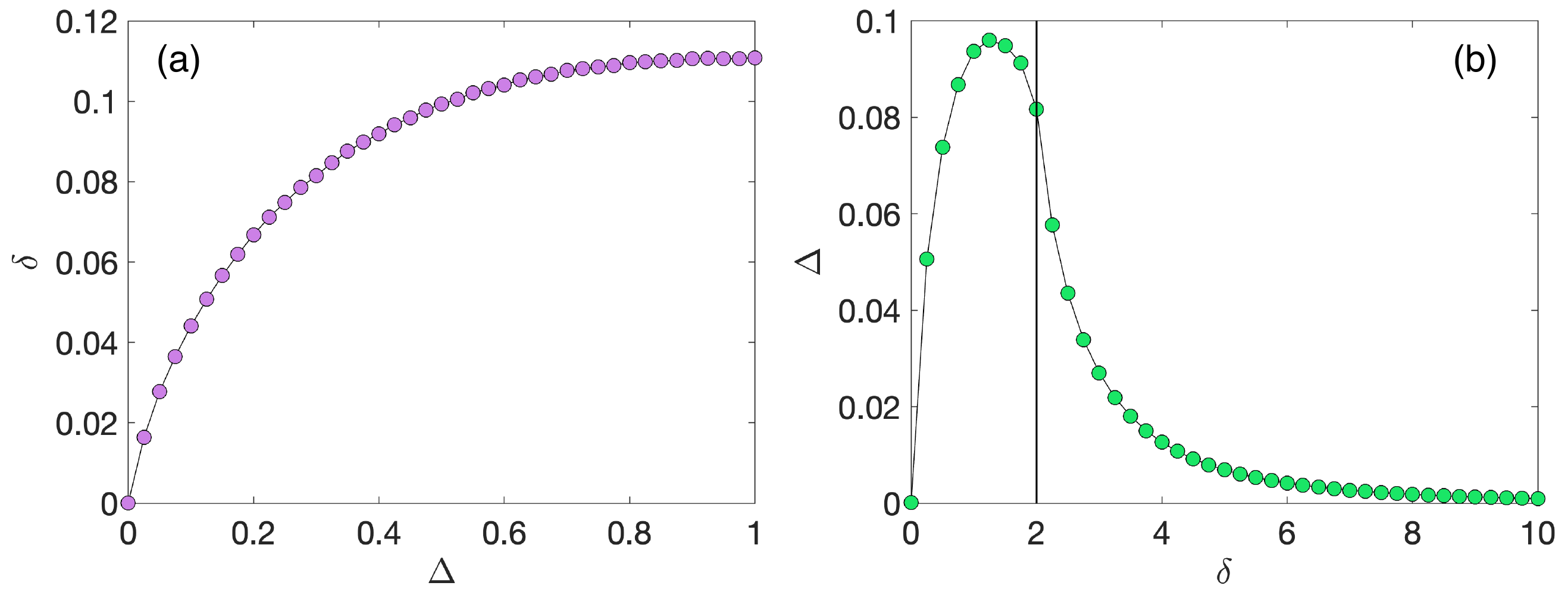}

		\caption{(a) The change of site nematic order when increasing NNN bond order into the meanfield hamiltonian. (b) The change of bond nematic order when increasing the site nematic order in the meanfield hamiltonian. The black line refers to where the Dirac points merge into each other and are about to gap out.}
		\label{mf_nematic_orders}
	\end{figure}
	
	\subsection{Section \uppercase\expandafter{\romannumeral9}: First-order transition between QAH and Stripe }
	To show more details of the transition between QAH and Stripe, we show the structure factors of the two phases at different temperature near their transition point in Fig.\ref{cut_aqh_stripe}. It is clear that near the first-order transition point $V_2\approx 2.6$, $J_\mathrm{QAH}$ suddenly drops while  $S_\mathrm{stripe}$ jumps to a high value from 0.
	\begin{figure}[H]
		\centering
		
		\includegraphics[width=0.4\textwidth]{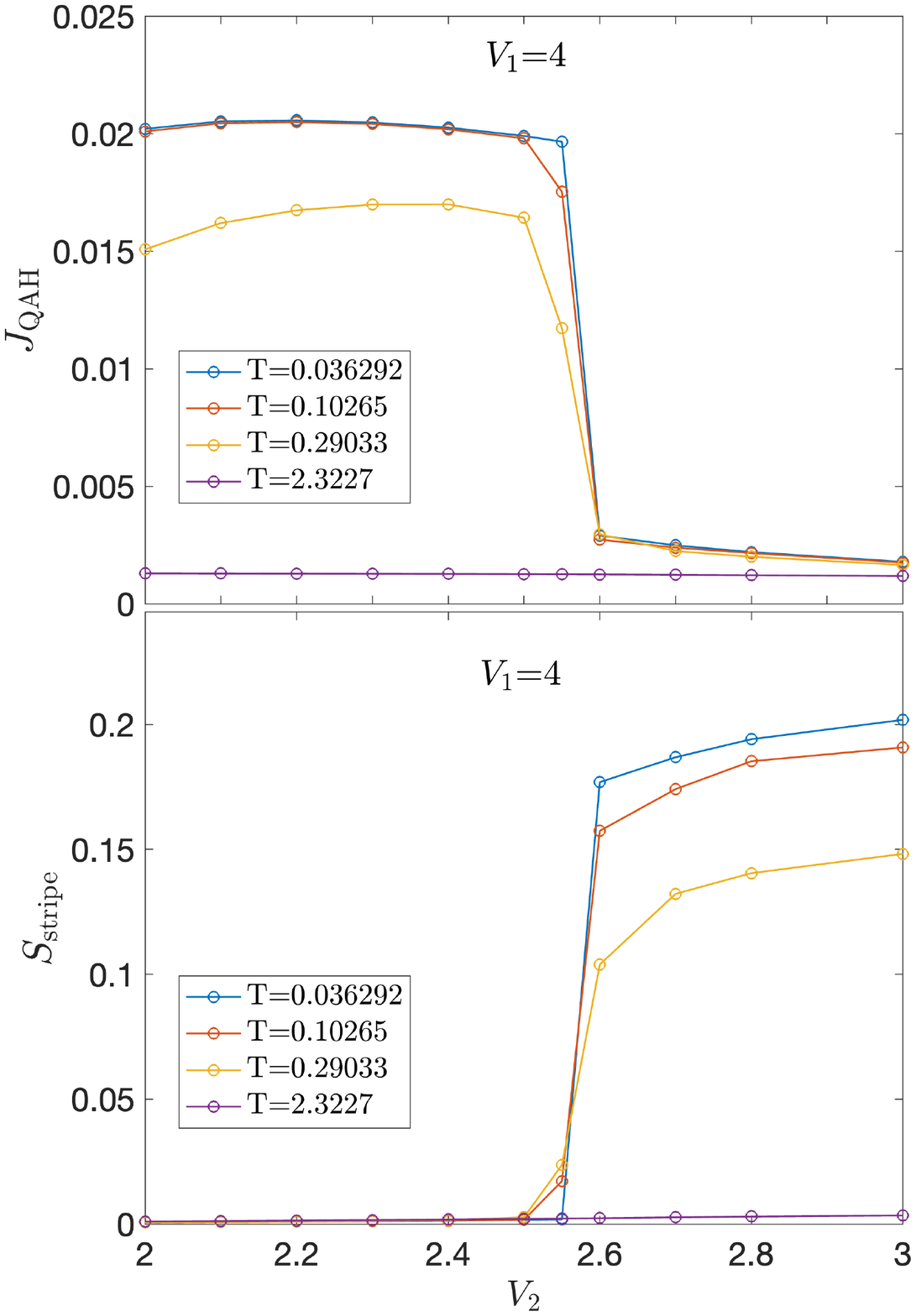}
		
		\caption{ With fixed $V_1=4$, the structure factors of QAH and stripe insulator with the change of $V_2$ at different temperture are shown.}
		\label{cut_aqh_stripe}
	\end{figure}
	
\end{widetext}

\end{document}